\documentclass[a4paper,fleqn]{cas-dc}

\usepackage[authoryear,longnamesfirst]{natbib}

\usepackage{amssymb}
\usepackage{amsmath}
\usepackage{float}
\usepackage{placeins}

\usepackage{ragged2e}

\usepackage{tabularx}
\usepackage{booktabs}
\usepackage{tabularx}
\usepackage{array}
\usepackage{longtable}
\usepackage{booktabs} 

\def\tsc#1{\csdef{#1}{\textsc{\lowercase{#1}}\xspace}}
\tsc{WGM}
\tsc{QE}
\tsc{EP}
\tsc{PMS}
\tsc{BEC}
\tsc{DE}

\begin{document}
\let\WriteBookmarks\relax
\def\floatpagepagefraction{1}
\def\textpagefraction{.001}

\shorttitle{Sensing-Aware Backscatter Communications}

\shortauthors{Rahul Gulia et~al.}


\title[mode = title]{\LARGE Sensing-Aware Backscatter Communications: A Survey on Envelope Stability, Waveform Design, and Selection Diversity}


\author[1]{Rahul Gulia\corref{cor1}}
\ead{rg9828@rit.edu}

\author[2]{Priyanka Mann}

\author[3]{Ashish Sheikh}

\author[4]{Feyisayo Favour Popoola}

\author[5]{Serisha Vadlamudi}


\affiliation[1]{
    organization={Rochester Institute of Technology},
    city={Rochester},
    state={NY},
    country={USA}
}

\affiliation[2]{
    organization={ Indian Institute of Technology},
    city={Delhi},
    state={New Delhi},
    country={India}
}

\affiliation[3]{
    organization={Cisco Systems Inc.},
    city={Richfield},
    state={OH},
    country={USA}
}

\affiliation[4]{
    organization={Obafemi Awolowo University},
    city={Ile-Ife},
    country={Nigeria}
}

\affiliation[5]{
    organization={The University of Texas at Dallas},
    city={Richardson},
    state={TX},
    country={USA}
}

\cortext[cor1]{Corresponding author}



\begin{abstract}
The vision of ubiquitous battery-free connectivity has positioned backscatter communication as a foundational technology for next-generation Internet of Things (IoT) and industrial sensing applications. However, as backscatter systems evolve from simple identification tags toward high-fidelity integrated sensing and communication (ISAC) devices, fundamental challenges emerge at the intersection of waveform design, hardware nonlinearity, and channel dynamics. This paper presents a comprehensive survey of sensing-aware backscatter communications, with a focus on three tightly coupled dimensions: envelope stability, spatial diversity, and temporal channel robustness. We introduce the concept of the ``Illuminator's Dilemma'' to describe the inherent conflict between high-peak-to-average power ratio (PAPR) multicarrier waveforms optimized for active communication and the stringent dynamic range requirements of passive tags. The survey provides a unified treatment of multi-objective waveform design strategies, a comprehensive taxonomy of near-far mitigation techniques, and an in-depth analysis of channel state information (CSI) aging in backscatter networks. Key contributions include tag-centric metrics such as the Backscatter Crest Factor (BCF), Envelope Stability Factor (ESF), and Sensing Fidelity Index (SFI), along with a comparative evaluation of PAPR reduction techniques, spatial diversity methods, and non-coherent detection schemes. This survey is intended for researchers, engineers, and graduate students working in wireless communications, IoT, RF sensing, and integrated sensing and communications.
\end{abstract}

\begin{keywords}
Backscatter communication \sep ambient IoT \sep waveform design \sep envelope stability \sep transmit antenna selection \sep channel state information aging \sep machine learning \sep integrated sensing and communication
\end{keywords}

\maketitle

\section{Introduction}

The proliferation of battery-free wireless systems has positioned backscatter communication as a foundational technology for next-generation Internet of Things (IoT), smart infrastructure, and industrial sensing applications \cite{vikram2023breaking, boyd2022backscatter, liu2022ambient, daskalakis2022rf}. By enabling ultra-low-power or passive devices to communicate through reflection rather than active radio frequency (RF) generation, backscatter systems eliminate the need for onboard power sources, thereby enabling scalable and maintenance-free deployments in environments such as smart warehouses, structural health monitoring systems, and logistics networks \cite{griffin2022multistatic, nikitin2022rfid, dobbins2022warehouse}.

Recent advances have extended the scope of backscatter beyond identification and low-rate data transmission toward integrated sensing and communication (ISAC) paradigms \cite{zhang2023isac, liu2023joint}. In particular, sensing-aware backscatter systems—such as surface acoustic wave (SAW)-based tags and RF-based environmental sensors—rely not only on reliable communication links but also on the preservation of signal fidelity at the tag interface \cite{reindl2021saw, hauptmann2022saw, galler2022wireless}. This introduces a fundamental shift in system design: unlike conventional wireless systems, where waveform optimization is primarily driven by spectral efficiency or bit error rate, backscatter systems must account for how the incident RF waveform interacts with nonlinear tag hardware and sensing mechanisms \cite{valenta2022rectenna, hemour2022nonlinear, boaventura2022waveform}.

A critical yet underexplored aspect of this interaction is the temporal envelope stability of the illuminating waveform. Modern wireless systems widely employ high peak-to-average power ratio (PAPR) multicarrier waveforms such as orthogonal frequency division multiplexing (OFDM) to maximize spectral efficiency \cite{hanzo2011ofdm, jiang2009papr, rahmatallah2013peak}. However, when such signals illuminate passive tags, the resulting envelope fluctuations can lead to severe degradation in both energy harvesting and sensing accuracy \cite{boaventura2022waveform, valenta2014harvesting, claveau2022waveform}. Specifically, envelope nulls may cause intermittent energy starvation, preventing reliable tag operation, while envelope peaks can drive rectifiers into nonlinear regimes, generating distortion and spectral artifacts \cite{boaventura2014boosting, kim2014wireless, boaventura2022waveform}. For sensing-enabled tags, these effects can manifest as resonance drift, missed echoes, or nonlinear mixing products, thereby directly impairing sensing fidelity \cite{reindl1998saw, bulst1999saw, plessky2021saw}.

This challenge gives rise to what we term the \textbf{``Illuminator's Dilemma''}: the need to balance high transmit power and waveform efficiency for the primary communication link with the stringent dynamic range constraints of passive backscatter devices. Unlike active receivers, passive tags lack adaptive gain control and are highly sensitive to instantaneous signal variations, making them particularly vulnerable to both saturation and starvation effects \cite{nikitin2010rfid, finkenzeller2010rfid, dobbins2021energy}. Despite its importance, this tradeoff has received limited attention in existing surveys, which predominantly focus on communication-centric metrics \cite{van2022survey, memon2022survey, khan2022backscatter}.

In parallel, large-scale deployments of backscatter systems in industrial environments introduce a pronounced near–far interference imbalance, exacerbated by the double path-loss characteristic of backscatter channels \cite{griffin2009multipath, kim2009range, boyer2012space}. Tags located close to the illuminator may experience saturation, while distant tags suffer from insufficient incident power, resulting in heterogeneous sensing and communication performance across the network \cite{arnitz2013multistatic, ma2020spatial, basharat2022irs}. While prior works have explored mitigation strategies such as power control, beamforming, intelligent reflecting surfaces (IRS), and multi-reader architectures \cite{nguyen2022mimo, wu2021irs, basar2020wireless, zamora2022cooperative}, a unified perspective that connects these approaches to sensing fidelity remains lacking.

Another fundamental challenge arises from the rapid temporal variation of backscatter channels, which leads to channel state information (CSI) aging \cite{li2020channel, ma2022channel, kim2021channel}. Unlike conventional systems with frequent pilot-based updates, passive tags often rely on blind or semi-blind estimation, making them particularly susceptible to outdated CSI \cite{zhang2019blind, qian2021noncoherent, guo2020differential}. This has direct implications for advanced techniques such as transmit antenna selection (TAS), where suboptimal decisions based on stale CSI can further exacerbate envelope instability and near–far effects \cite{chen2022antenna, vicario2022antenna, serisha2022tas}.

Motivated by these challenges, this paper presents a comprehensive survey of sensing-aware backscatter communication systems, with a focus on three tightly coupled dimensions: (i) envelope stability and waveform design, (ii) spatial diversity and selection mechanisms, and (iii) temporal dynamics and CSI robustness. In contrast to existing surveys, we adopt a tag-centric perspective, emphasizing how system-level design choices impact both communication reliability and sensing fidelity.

Finally, we identify promising future research directions, including the potential of machine learning frameworks for joint waveform-spatial-temporal optimization, which we discuss in Section~\ref{sec:conclusion_future}.

The main contributions of this survey are summarized as follows:

\begin{itemize}
    \item We \textbf{formalize envelope stability} as a central design dimension for sensing-aware backscatter systems and introduce tag-centric performance metrics, including the Backscatter Crest Factor (BCF), Envelope Stability Factor (ESF), and Sensing Fidelity Index (SFI), to quantify its impact on energy harvesting and sensing fidelity.

    \item We provide a \textbf{unified review of multi-objective waveform design strategies}, bridging insights from PAPR reduction, wireless power transfer (WPT), and backscatter communication to address the ``Illuminator's Dilemma.''

    \item We present a \textbf{taxonomy of near-far mitigation techniques}, including spatial, waveform, and protocol-level approaches, and provide a comparative evaluation of their suitability for passive sensing applications.

    \item We examine the \textbf{effects of CSI aging and non-coherent detection} in backscatter networks, highlighting their implications for antenna selection, channel tracking, and system robustness in mobile scenarios.

    \item We \textbf{situate the present work within the broader literature} through a comprehensive review of existing surveys, identifying gaps in communication-centric backscatter surveys, WPT-focused waveform design surveys, and ISAC surveys that overlook the unique challenges of passive, tag-centric sensing.

    \item We identify \textbf{promising future research directions}, including machine learning frameworks for joint optimization, ISAC co-design, RIS-assisted envelope shaping, digital twin-assisted planning, and 6G Ambient IoT ecosystems.
\end{itemize}

The remainder of this paper is organized as follows. Section II provides the foundational system model and introduces key performance metrics for sensing-aware backscatter design. Section III reviews related surveys and positions this work within the existing literature. Section IV delves into the envelope stability challenge, presenting a comparative analysis of waveforms and their impact on tag operation. Section V addresses the near-far interference gap through a comprehensive review of spatial mitigation techniques. Section VI tackles CSI aging and non-coherent detection methods. Finally, Section VII discusses open challenges, future research directions, and concludes the paper.

\section{Foundations of Sensing-Aware Backscatter Systems}

Before delving into the specific challenges of envelope stability, spatial diversity, and temporal dynamics, it is essential to establish the foundational system model and performance metrics that underpin sensing-aware backscatter communications. This section provides the analytical framework that will be used throughout the survey, formalizes tag-centric metrics that quantify sensing fidelity, and presents a taxonomy that organizes the subsequent technical sections.

\subsection{System Model and Architecture}

A canonical backscatter system consists of three fundamental entities: an \textit{illuminator} that generates the incident RF waveform, one or more \textit{passive tags} that modulate and reflect this incident signal, and a \textit{reader} that demodulates the backscattered information \cite{boyd2022backscatter, griffin2022multistatic, tornatta2022backscatter}. Unlike conventional active communications, where the transmitter and receiver are symmetric in their hardware capabilities, backscatter links exhibit a fundamental asymmetry: the illuminator may be a high-power dedicated carrier emitter or an ambient source such as a WiFi access point, while the tag is an ultra-low-power device with no active RF generation capability \cite{liu2022ambient, dunna2022ambient, bharadia2022full, ahmed2023backscatter}.

\begin{figure*}[H]
\centering
\includegraphics[width=0.99\textwidth]{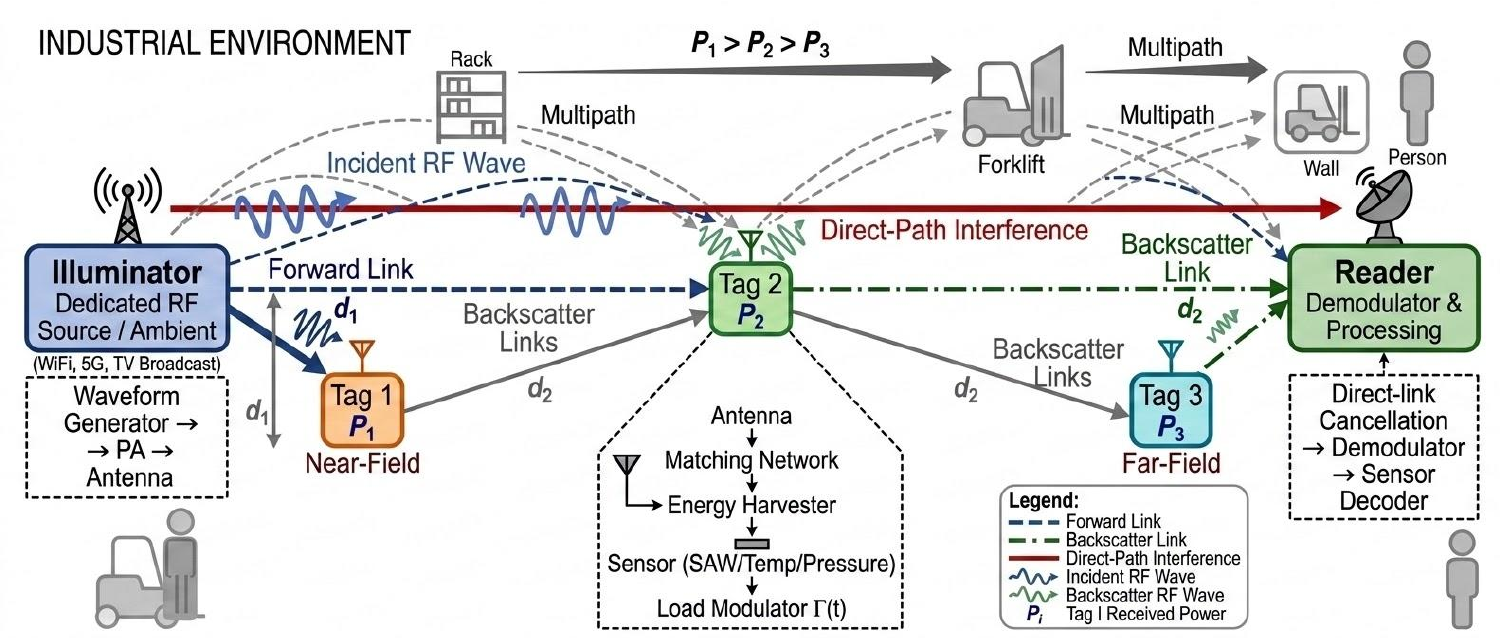}
\caption{Canonical sensing-aware backscatter communication architecture showing the illuminator, three passive tags at varying distances, and the reader. The power gradient $P_1 > P_2 > P_3$ illustrates the effect of double path loss. Tag 2 is shown with an exploded internal architecture. Forward, backscatter, direct-path, and multipath signal components are indicated.}
\label{fig:system_architecture}
\end{figure*}

The signal received at the reader consists of two primary components: the direct path from the illuminator and the backscattered path from the tag, as shown in Figure~\ref{fig:system_architecture}. This fundamental relationship is modeled as:

\begin{equation}
y(t) = \underbrace{h_d x(t)}_{\text{direct link}} + \underbrace{h_{\text{casc}} \Gamma(t) x(t)}_{\text{backscatter link}} + n(t)
\label{eq:received}
\end{equation}

where $h_{\text{casc}} = h_f h_b$ is the cascaded backscatter channel, comprising the forward link $h_f$ (illuminator $\rightarrow$ tag) and the backward link $h_b$ (tag $\rightarrow$ reader), $h_d$ represents the direct channel coefficient, $\Gamma(t)$ is the tag's time-varying reflection coefficient, and $n(t)$ is additive noise. The direct link component $h_d x(t)$ is typically orders of magnitude stronger than the backscatter component, creating the fundamental challenge of self-interference that all backscatter systems must address \cite{nikitin2010rfid, bharadia2022full, kim2023direct}. This interference is particularly severe for distant tags where $|h_{\text{casc}}| \ll |h_d|$, directly motivating the near-far mitigation techniques surveyed in Section IV.

The end-to-end backscatter channel experiences \textit{double path loss}, which can be expressed as:

\begin{equation}
\begin{aligned}
P_{\mathrm{reader}}
&= P_{\mathrm{illuminator}}
   G_{\mathrm{ill}}
   G_{\mathrm{tag}}
   G_{\mathrm{reader}} \\
&\quad \times
\left(\frac{\lambda}{4\pi d_1}\right)^2
\left(\frac{\lambda}{4\pi d_2}\right)^2
|\Gamma|^2
|H|^2
\end{aligned}
\label{eq:pathloss}
\end{equation}

where $d_1$ is the illuminator-tag distance, $d_2$ is the tag-reader distance, $\Gamma$ is the tag's modulation reflection coefficient, and $H$ captures multipath fading effects \cite{griffin2009multipath, kim2009range, boyer2012space, wang2024ambient}. This double path loss means that the received power decays as $(d_1 d_2)^{-2}$, making the near-far disparity significantly more severe than in conventional point-to-point links \cite{nikitin2006path, lazaro2009analysis}.

For sensing-aware backscatter, the tag incorporates not only a communication modulator but also a sensing element. Surface acoustic wave (SAW) sensors represent a prominent example, where the backscattered signal contains echoes whose time delays and phase shifts encode physical parameters such as temperature, strain, or pressure \cite{reindl1998saw, reindl2021saw, hauptmann2022saw, plessky2021saw, rodriguez2024passive}. In such systems, the incident waveform must simultaneously:
\begin{itemize}
    \item Provide sufficient energy to activate the tag's circuitry,
    \item Enable modulation of identification or sensor data, and
    \item Preserve the fidelity of the sensing information embedded in the backscattered signal.
\end{itemize}

The equivalent complex baseband signal at the tag is:

\begin{equation}
s(t) = \sum_{n=0}^{N-1} X_n e^{j2\pi f_n t}
\label{eq:complex_envelope}
\end{equation}

where $X_n$ is the complex symbol on the $n$-th subcarrier. The corresponding RF waveform is $x_{\mathrm{RF}}(t) = \Re\{ s(t) e^{j2\pi f_c t} \}$. Throughout this paper, envelope statistics (PAPR, PMR, BCF, ESF) are evaluated using the equivalent complex baseband representation $|s(t)|$, following standard OFDM analysis.

where $a_n(t)$ represents the complex envelope of the $n$-th subcarrier. For multi-carrier waveforms such as OFDM, these envelopes exhibit significant fluctuations that directly impact tag operation \cite{hanzo2011ofdm, jiang2009papr, chen2023sensing}.

\subsection{The Nonlinear Tag: Rectifier and Sensor Models}

\begin{figure*}[H]
\centering
\includegraphics[width=0.8\textwidth, height=0.25\textwidth]{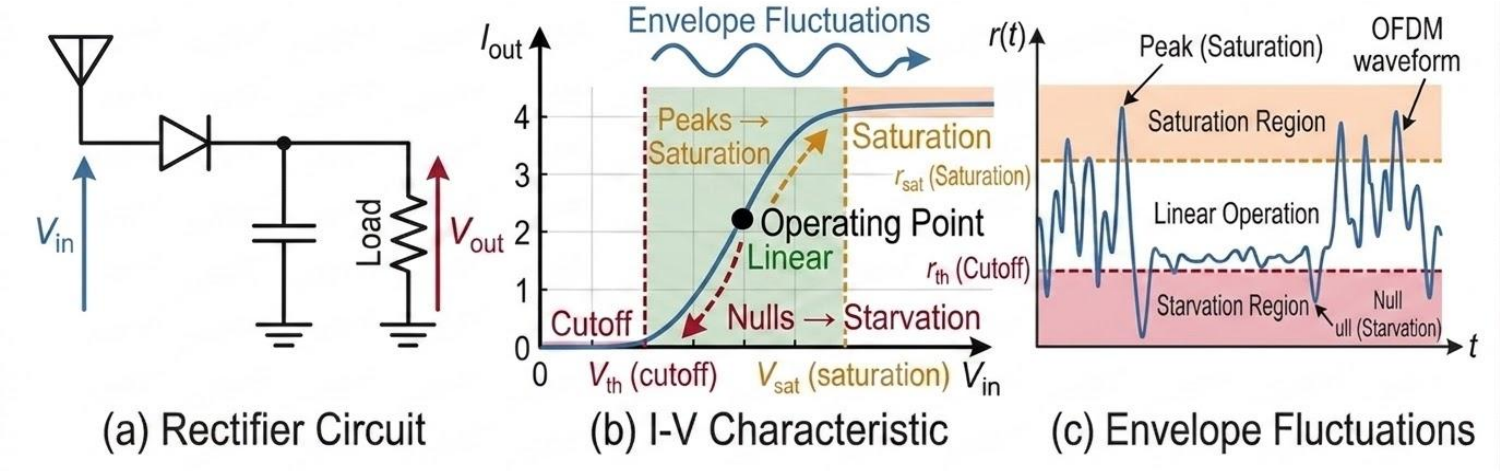}
\caption{Rectifier nonlinear characteristics showing operating regions and the effect of envelope fluctuations on operating point migration. The inset demonstrates how envelope peaks drive the rectifier into saturation while nulls cause cutoff.}
\label{fig:rectifier}
\end{figure*}

The passive tag's front-end consists of a rectifier circuit that converts incident RF power to DC, enabling tag operation, as shown in Figure~\ref{fig:rectifier}. This rectifier exhibits inherently nonlinear behavior that is fundamental to understanding envelope stability \cite{valenta2022rectenna, hemour2022nonlinear, boaventura2022waveform, clerckx2023waveform}.

The rectifier's output voltage $V_{\mathrm{out}}$ is a nonlinear function of the input RF voltage $V_{\mathrm{in}}$:

\begin{equation}
V_{\mathrm{out}} = f(V_{\mathrm{in}}) \approx \sum_{k=0}^{\infty} \beta_k V_{\mathrm{in}}^k
\label{eq:nonlinear}
\end{equation}

where $\beta_k$ are device-dependent coefficients \cite{boaventura2022waveform, clerckx2021fundamentals, kim2024machine}. The instantaneous power conversion efficiency $\eta(t)$ depends critically on the input signal envelope:

\begin{equation}
\eta = \frac{\frac{1}{T}\int_0^T P_{\mathrm{DC}}(t)\,dt}{\frac{1}{T}\int_0^T P_{\mathrm{RF}}(t)\,dt}
\label{eq:efficiency}
\end{equation}

The average power conversion efficiency $\eta$ is a nonlinear function of the incident waveform envelope, depending on its statistical distribution, peak-to-average ratio, higher-order moments, load conditions, and rectifier operating point \cite{clerckx2021fundamentals, valenta2022rectenna, hemour2022nonlinear}.

For sensing-enabled tags, envelope instability introduces additional degradation mechanisms. In SAW sensors, the reflected signal consists of multiple echoes:

\begin{equation}
y_{\mathrm{saw}}(t) = \sum_{m=1}^{M} \alpha_m e^{j\phi_m} x(t - \tau_m)
\label{eq:saw}
\end{equation}

where $\alpha_m$, $\phi_m$, and $\tau_m$ encode sensing information \cite{reindl1998saw, bulst1999saw}. Envelope fluctuations introduce phase noise and amplitude modulation that corrupt these parameters, directly impairing sensing fidelity \cite{galler2022wireless, hauptmann2022saw, chen2024sensing}.

\subsection{Tag-Centric Performance Metrics}
\label{sec:metrics}

To systematically evaluate sensing-aware backscatter systems, this survey formalizes a set of tag-centric metrics that capture the unique requirements of passive sensing. These metrics will be used throughout the survey to compare different approaches.

\begin{table*}[t]
\centering
\caption{Tag-Centric Performance Metrics for Sensing-Aware Backscatter}
\label{tab:metrics}
\footnotesize
\renewcommand{\arraystretch}{1.3}
\setlength{\tabcolsep}{11pt}
\begin{tabular}{|p{2.5cm}|p{0.7cm}|p{4.6cm}|p{6.5cm}|}
\hline
\textbf{Metric} & \textbf{Symbol} & \textbf{Definition} & \textbf{Why It Matters} \\
\hline
Envelope Stability Factor & $ESF$ & $\displaystyle \frac{\min |x(t)|^2}{\max |x(t)|^2}$ & Quantifies incident waveform flatness; inversely related to PAPR; determines harvesting efficiency and sensing fidelity \\
\hline
Sensing-to-Clutter Ratio & $SCR$ & $\displaystyle \frac{P_{\mathrm{sensing}}}{P_{\mathrm{clutter}}} = \frac{|\sum_{m=1}^M \alpha_m e^{j\phi_m}|^2 \cdot \mathbb{E}[|x(t)|^2]}{\mathbb{E}[|h_d x(t) + n(t)|^2]}$ & Measures sensing signal power relative to direct path interference and noise; aligned with radar literature \\
\hline
Harvesting Outage Probability & $P_{\mathrm{out}}$ & $\displaystyle \Pr\left(\int_0^T \eta(t)P_{\mathrm{RF}}(t)dt < E_{\mathrm{th}}\right)$ & Probability that harvested energy falls below required threshold over operation period \\
\hline
Sensing Fidelity Index & $SFI$ & 
$\displaystyle \frac{1}{1 + \tilde{\sigma}_{\phi}^2 + \tilde{\sigma}_{\alpha}^2}$
& Quantifies sensing degradation due to envelope-induced phase and amplitude noise \\
\hline
Backscatter Modulation Efficiency & $BME$ & $\displaystyle \frac{P_{\mathrm{mod}}}{P_{\mathrm{inc}}} = |\Delta\Gamma|^2$ & Fraction of incident power converted to modulated backscatter \\
\hline
Channel Coherence Time & $T_c$ & $\displaystyle \int_0^{\infty} |\rho(\Delta t)|^2 d(\Delta t)$ & Timescale over which channel remains correlated; critical for CSI aging analysis \\
\hline
\end{tabular}
\end{table*}

The \textbf{Envelope Stability Factor (ESF)} quantifies the flatness of the incident waveform envelope. For an ideal continuous wave (CW) signal, $ESF = 1$. For high-PAPR waveforms such as OFDM, $ESF \ll 1$, indicating severe envelope fluctuations. 

Unlike PMR, which increases as envelope fluctuations worsen, ESF is bounded between 0 and 1, with larger values indicating greater envelope stability. This normalization makes ESF particularly convenient for comparing waveform suitability across different sensing-aware backscatter scenarios.

\begin{equation}
\mathrm{ESF} = \frac{1}{\mathrm{PMR}} = \frac{\min |x(t)|^2}{\max |x(t)|^2},
\label{eq:esf_pmr}
\end{equation}

where $\mathrm{PMR} = \max |x(t)|^2 / \min |x(t)|^2$ \cite{rettelbach2012pmr}. Unlike the conventional peak-to-average power ratio (PAPR), which measures peak power relative to the average signal power, PMR additionally captures the severity of envelope nulls. Since passive backscatter tags are sensitive to both excessive envelope peaks and deep fades, PMR (or equivalently ESF) provides a useful complementary metric for evaluating envelope stability.

To quantify the impact of envelope instability on sensing performance, this survey introduces the Sensing Fidelity Index (SFI), a dimensionless metric defined as:

\begin{equation}
SFI = \frac{1}{1 + \tilde{\sigma}_{\phi}^2 + \tilde{\sigma}_{\alpha}^2}
\label{eq:sfi}
\end{equation}

where $\tilde{\sigma}_{\phi}^2$ and $\tilde{\sigma}_{\alpha}^2$ denote the normalized variances of phase and amplitude fluctuations, respectively, induced by envelope instability. The variances are normalized with respect to their maximum allowable values for reliable SAW sensing, ensuring SFI is dimensionless and bounded between 0 and 1. The inverse form ensures that SFI decreases monotonically with increasing phase and amplitude instability, with $SFI = 1$ indicating perfect sensing fidelity and $SFI \to 0$ under severe degradation. Inspired by the sensitivity of SAW sensing to phase and amplitude fluctuations reported in \cite{reindl2021saw, plessky2021saw, zhang2024noncoherent}, this metric provides a unified measure of sensing fidelity under envelope instability.

\begin{table}[t]
\centering
\caption{Mapping of Key Metrics to Technical Sections}
\label{tab:metric_mapping}
\footnotesize
\renewcommand{\arraystretch}{1.2}
\begin{tabularx}{\columnwidth}{|p{2.5cm}|p{1cm}|X|}
\hline
\textbf{Metric} & \textbf{Section} & \textbf{Role} \\
\hline
Envelope Stability Factor (ESF) & III & Quantifies waveform--tag interaction. \\
\hline
Path Loss Ratio ($\rho = d_1 d_2 / \lambda^2$) & IV & Measures spatial disparity. \\
\hline
Channel Coherence Time ($T_c$) & V & Determines CSI tracking requirements. \\
\hline
Sensing Fidelity Index (SFI) & III--V & Unified sensing performance metric. \\
\hline
\end{tabularx}
\end{table}


\subsection{Taxonomy of Sensing-Aware Design Dimensions}
\label{sec:taxonomy}

\begin{figure}
\centering
\includegraphics[width=0.5\textwidth]{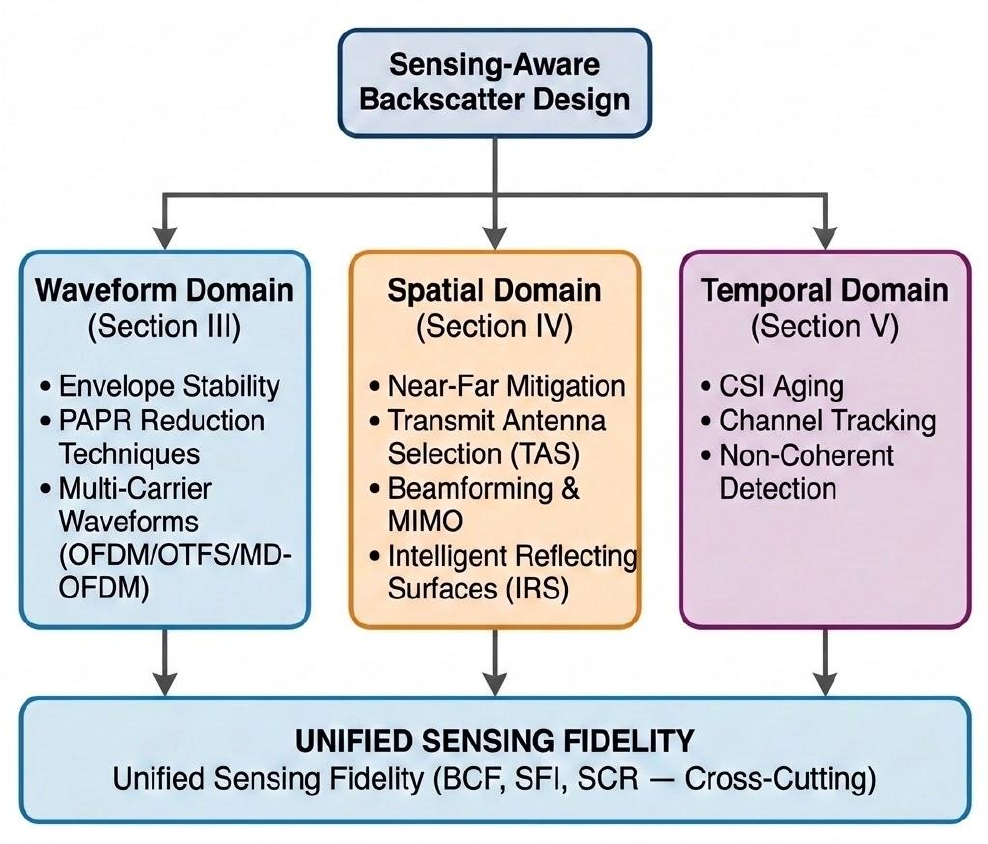}
\caption{Taxonomy of sensing-aware backscatter design dimensions. The hierarchical structure organizes the survey's technical sections and illustrates relationships among different solution approaches.}
\label{fig:taxonomy}
\end{figure}

Based on the system model and metrics developed above, the sensing-aware backscatter design space is organized into three fundamental dimensions, each addressed in subsequent sections. Figure~\ref{fig:taxonomy} presents this taxonomy visually.

The three primary dimensions are:

\begin{itemize}
    \item \textbf{Waveform Domain (Section III)}: Addresses the ``Illuminator's Dilemma'' through analysis of how primary waveform characteristics—particularly envelope stability—affect tag harvesting and sensing. This includes review of PAPR reduction techniques, multi-sine waveform design for WPT, and comparative analysis of OFDM, OTFS, and MD-OFDM \cite{boaventura2022waveform, ma2021waveform, li2024isac}.
    
    \item \textbf{Spatial Domain (Section IV)}: Tackles the near-far interference gap through spatial diversity techniques. Approaches include transmit antenna selection (TAS), beamforming, intelligent reflecting surfaces (IRS), and cooperative tag relaying \cite{chen2022antenna, basharat2022irs, zamora2022cooperative, kumar2023spatial}.
    
    \item \textbf{Temporal Domain (Section V)}: Addresses CSI aging and channel dynamics in mobile backscatter scenarios. This encompasses channel estimation techniques, Kalman filtering, non-coherent detection, and angle-of-arrival (AoA) based methods \cite{li2020channel, ma2022channel, qian2021noncoherent, kumar2023channel, zhang2024noncoherent}.
\end{itemize}

These three dimensions—waveform, spatial, and temporal—are intrinsically coupled: improving transmit power to enhance sensing for distant tags may drive near-field tags into saturation, while waveform optimization for envelope stability may reduce spectral efficiency. This interdependence motivates the unified sensing fidelity framework (BCF, SFI, SCR) discussed throughout this survey and the future research directions outlined in Section~\ref{sec:conclusion}.

\subsection{Comparison with Existing Surveys}
\label{sec:comparison}

Table~\ref{tab:comparison} positions this survey relative to existing comprehensive reviews in the field, highlighting the unique contributions of the sensing-aware, tag-centric perspective adopted in this work.

\begin{table}[t]
\centering
\caption{Comparison of Existing Backscatter Surveys with This Work}
\label{tab:comparison}
\footnotesize
\setlength{\tabcolsep}{4pt}

\begin{tabular}{p{3.0cm} c c c c c c}
\toprule
\textbf{Survey} &
\textbf{Year} &
\textbf{WF} &
\textbf{ES} &
\textbf{NF} &
\textbf{CSI} &
\textbf{SF} \\
\midrule

\cite{van2022survey} & 2025 & $\circ$ & $\circ$ & $\bullet$ & $\circ$ & $\circ$ \\
\cite{memon2022survey} & 2023 & $\bullet$ & $\circ$ & $\bullet$ & $\circ$ & $\circ$ \\
\cite{khan2022backscatter} & 2023 & $\bullet$ & $\circ$ & $\bullet$ & $\circ$ & $\circ$ \\
\cite{basharat2022irs} & 2022 & $\circ$ & $\circ$ & $\bullet$ & $\circ$ & $\circ$ \\
\cite{vikram2023breaking} & 2024 & $\bullet$ & $\circ$ & $\bullet$ & $\bullet$ & $\circ$ \\
\cite{ahmed2023backscatter} & 2026 & $\bullet$ & $\circ$ & $\bullet$ & $\circ$ & $\circ$ \\
\cite{li2024isac} & 2023 & $\bullet$ & $\circ$ & $\circ$ & $\circ$ & $\bullet$ \\
\cite{wang2024ambient} & 2020 & $\bullet$ & $\circ$ & $\bullet$ & $\bullet$ & $\circ$ \\
\cite{chen2025backscatter} & 2025 & $\bullet$ & $\circ$ & $\bullet$ & $\bullet$ & $\circ$ \\
\cite{kumar2025isac} & 2025 & $\bullet$ & $\circ$ & $\circ$ & $\circ$ & $\bullet$ \\
\cite{rodriguez2025ambient} & 2025 & $\bullet$ & $\circ$ & $\bullet$ & $\bullet$ & $\circ$ \\

\midrule
\textbf{This Survey} & \textbf{2026} &
\textbf{$\bullet$} &
\textbf{$\bullet$} &
\textbf{$\bullet$} &
\textbf{$\bullet$} &
\textbf{$\bullet$} \\
\bottomrule
\end{tabular}

\vspace{2mm}
\footnotesize
\textbf{Abbreviations:} WF = Waveform Design, ES = Envelope Stability, NF = Near--Far Mitigation,
CSI = CSI Aging, SF = Sensing Focus. \\
$\bullet$ = Comprehensive coverage; $\circ$ = Limited or no coverage.

\end{table}

As Table~\ref{tab:comparison} demonstrates, existing surveys have primarily focused on communication-centric metrics and have not systematically addressed the interaction between waveform envelope characteristics and sensing fidelity. The concept of envelope stability as a unified framework linking waveform design, tag hardware constraints, and sensing performance represents the primary novelty of this work.

\subsection{Design Implications and Chapter Roadmap}
\label{sec:roadmap}

The metrics and models developed in this section provide a unified framework for understanding sensing-aware backscatter as a multi-objective optimization problem. From a system-level perspective, the design of sensing-aware backscatter represents a joint optimization across coupled dimensions. The Envelope Stability Factor (ESF) directly couples waveform design (Section III) to tag harvesting and sensing fidelity. The double path-loss ratio $\rho = d_1 d_2 / \lambda^2$ governs the spatial disparity addressed in Section IV. The channel coherence time $T_c$ limits the validity of CSI, motivating the temporal tracking and non-coherent techniques reviewed in Section V. 

These three dimensions—waveform, spatial, and temporal—are intrinsically coupled: improving transmit power to enhance sensing for distant tags may drive near-field tags into saturation, while waveform optimization for envelope stability may reduce spectral efficiency. This interdependence motivates the emerging machine learning frameworks explored in Section VI, which enable joint optimization across all dimensions. Figure~\ref{fig:taxonomy} visually summarizes this organization, which structures the remainder of this survey.

\section{Related Surveys}
\label{sec:related_work}

A number of surveys have reviewed different aspects of backscatter communication, ambient IoT, and integrated sensing and communication (ISAC). However, to the best of our knowledge, existing surveys do not jointly and systematically address the three core challenges that define sensing-aware backscatter systems: envelope stability, near-far interference, and CSI aging. This section situates the present work within the broader literature and highlights the gaps it fills.

\subsection{Surveys on Backscatter and Ambient IoT}
\label{sec:related_backscatter}

Several surveys provide broad overviews of backscatter communication, focusing primarily on system architectures, modulation schemes, and application domains.

\cite{van2022survey} review ambient backscatter fundamentals, modulation, and networking protocols, but treat the incident waveform as a static excitation source and do not analyze envelope dynamics or rectifier constraints. \cite{memon2022survey} and \cite{khan2022backscatter} survey backscatter architectures and IoT applications, but their performance discussions center on communication-centric metrics such as BER, throughput, and coverage. \cite{liu2022ambient} survey ambient IoT and RF-powered systems, emphasizing energy harvesting and protocol design, but without addressing waveform-tag interactions or sensing fidelity.

Earlier foundational surveys on RFID and backscatter \cite{dobkin2012rfid, nikitin2010rfid, finkenzeller2010rfid} established the basic principles of passive tag operation, including rectifier design, impedance matching, and modulation schemes. However, these works focus primarily on identification and inventory management rather than sensing applications. More recent surveys on ambient backscatter \cite{kim2018ambient, ma2021backscatter, iyer2018inter-technology} discuss energy harvesting and communication protocols, but they model the incident signal as a quasi-stationary excitation source and do not consider the impact of waveform envelope fluctuations on tag operation.

Similarly, surveys on RF-powered systems and Internet of Things (IoT) applications \cite{lu2018wireless, wang2017ambient} emphasize energy efficiency and coverage but treat the illuminator waveform as a given rather than a design variable. They do not analyze how waveform characteristics affect the nonlinear dynamics of rectifier circuits or the fidelity of sensing measurements.

\textbf{Gap:} These surveys assume quasi-static illumination and do not consider how waveform envelope fluctuations, PA nonlinearity, or rectifier dynamics affect sensing-enabled tags. None introduce tag-centric metrics such as BCF, ESF, or SFI.

\subsection{Surveys on Wireless Power Transfer and Waveform Design}
\label{sec:related_wpt}

A separate line of surveys and research papers focuses on RF energy harvesting and waveform optimization. \cite{clerckx2019waveform, clerckx2021fundamentals, clerckx2017wireless, clerckx2018wireless} provide comprehensive treatments of waveform design for wireless power transfer (WPT), highlighting the role of multisine waveforms and rectifier nonlinearity in achieving high conversion efficiency. \cite{boaventura2022waveform, boaventura2016waveform} and \cite{collado2012waveform} survey WPT-optimized waveforms and their impact on rectifier operation. These works demonstrate that the instantaneous waveform shape significantly affects the efficiency of RF-to-DC conversion, particularly when rectifiers operate in nonlinear regimes.

\cite{rahmatallah2013peak} and \cite{jiang2009papr} survey PAPR reduction techniques for OFDM, but from a transmitter-centric viewpoint focused on power amplifier (PA) efficiency and spectral regrowth. \cite{wang2014papr} provide a comprehensive survey of PAPR reduction techniques, categorizing them into distortion-based, probabilistic, and structured methods. \cite{muller1997papr} and \cite{bauml1996papr} established the foundational principles of selected mapping (SLM) and partial transmit sequence (PTS), while \cite{tellambura2000papr} introduced active constellation extension (ACE).

More recently, surveys on simultaneous wireless information and power transfer (SWIPT) \cite{lu2018wireless, kim2015wireless} examine the tradeoff between information rate and harvested energy. However, these works focus on active receivers that can perform power splitting and do not consider passive backscatter tags that rely entirely on reflected signals.

\textbf{Gap:} These works do not connect waveform design to sensing fidelity, nor do they analyze envelope stability at the tag under cascaded backscatter channels. They also do not address spatial or temporal coupling (near-far imbalance, CSI aging). The tag-side quality of the incident waveform—as quantified by metrics such as ESF and BCF—is largely ignored.

\subsection{Surveys on Spatial Diversity, IRS, and MIMO Backscatter}
\label{sec:related_spatial}

Recent surveys explore spatial techniques for improving backscatter performance. IRS surveys \cite{basar2020wireless, wu2021irs, basharat2022irs, zhang2019irs, wu2019irs} discuss reconfigurable intelligent surface (RIS)-assisted backscatter and passive beamforming. These works show that IRS can significantly improve coverage and energy efficiency by controlling reflected paths. \cite{huang2019irs} and \cite{liang2021irs} provide comprehensive surveys of IRS architectures and optimization frameworks, while \cite{cao2021irs} focus on IRS-assisted backscatter communication specifically.

MIMO backscatter surveys \cite{nguyen2022mimo} review multi-antenna readers, spatial modulation, and diversity gains. \cite{sanayei2007mimo} provide a foundational survey of antenna selection techniques for MIMO systems, establishing the theoretical framework that underpins TAS. \cite{gorokhov2002antenna} analyze the performance of antenna selection in fading channels.

Cooperative backscatter surveys \cite{zamora2022cooperative, liu2013ambient} examine tag-to-tag relaying and multistatic architectures. These works show that cooperative strategies can extend coverage and improve reliability in dense deployments.\cite{chen2020irs} survey IRS-assisted backscatter communications specifically, highlighting the potential of RIS to enable passive beamforming for weak backscatter signals.

\textbf{Gap:} These surveys treat spatial techniques in isolation and do not analyze how spatial diversity interacts with rectifier saturation, envelope ripple, or tag dynamic range. None quantify the near-far gap using metrics like $\Gamma_{\mathrm{NF}}$, nor do they connect spatial mitigation to waveform or temporal dynamics. The interaction between spatial diversity and envelope stability—for example, how TAS can flatten the envelope at the tag—is not addressed.

\subsection{Surveys on Channel Estimation and CSI in Backscatter}
\label{sec:related_csi}

A smaller set of works focuses on channel estimation for passive systems. \cite{li2020channel} provide a survey of channel estimation techniques for backscatter communications, covering pilot-based, blind, and semi-blind approaches. \cite{truong2013channel} and \cite{varan2018channel} analyze channel estimation for ambient backscatter systems, highlighting the challenges of cascaded channel estimation and direct-path interference.

\cite{liu2021channel} survey channel estimation for backscatter communication systems, focusing on the tradeoff between estimation accuracy and training overhead. \cite{ma2022channel, ma2023channel} provide surveys of channel aging in wireless communication systems, but from the perspective of active networks rather than passive backscatter.

Surveys on ambient backscatter CSI \cite{zhang2019blind, qian2021noncoherent, guo2020differential} discuss cascaded channel estimation, pilot design, and non-coherent detection. Some works analyze tag selection and reader-side CSI proxies \cite{chen2022antenna, vicario2022antenna}. However, these surveys focus primarily on estimation algorithms rather than the temporal dynamics of the channel.

\textbf{Gap:} To the best of our knowledge, existing surveys do not jointly address CSI aging, temporal correlation, or the faster decorrelation of cascaded backscatter channels in the context of sensing-aware operation. The connection between CSI aging and envelope stability or near-far mitigation remains underexplored, and temporal metrics such as $T_c$ or $\rho(\Delta t)$ are not systematically introduced for backscatter systems. Furthermore, the coupling between CSI aging and spatial selection (e.g., TAS) has received limited attention in the existing literature.

\subsection{Surveys on ISAC and Joint Sensing-Communication Waveforms}
\label{sec:related_isac}

ISAC surveys \cite{liu2020isac, zhang2021isac, liu2022isac, wei2022isac} review joint radar-communication waveforms, sensing metrics, and optimization frameworks. \cite{liu2020isac} provide a comprehensive survey of integrated sensing and communication, covering waveform design, signal processing, and system architectures. \cite{zhang2021isac} and \cite{wei2022isac} focus on ISAC for 6G networks, highlighting the role of OFDM and massive MIMO.

These surveys emphasize active transceivers with high-power transmitters and sophisticated signal processing capabilities. They consider sensing applications such as radar, localization, and environmental mapping, where the transmitter actively emits signals and processes their reflections.

\textbf{Gap:} While these surveys provide valuable insights for active transceivers, they offer limited coverage of passive tags, rectifier constraints, and envelope stability. The unique challenges of battery-free sensing, double path loss, and cascaded channel aging are not systematically addressed in the existing literature. Moreover, waveform design for passive tag illumination—as distinct from active sensing—remains an area that has received limited attention

\subsection{Positioning of This Work}
\label{sec:related_positioning}

In contrast to the existing surveys summarized above, this work provides a unified, tag-centric perspective on sensing-aware backscatter communications. Table~\ref{tab:comparison} provides a detailed comparison of this survey with existing reviews.

Our contributions are threefold:

First, we formalize \textbf{envelope stability} through tag-centric metrics including the Backscatter Crest Factor (BCF), Envelope Stability Factor (ESF), and Sensing Fidelity Index (SFI). We re-evaluate classical PAPR reduction techniques through a sensing-aware lens and identify that no existing technique simultaneously controls peaks, nulls, and ripple—the three degradation channels for passive sensing.

Second, we quantify the \textbf{near-far gap} through the power ratio $\Gamma_{\mathrm{NF}}$ and provide a comprehensive taxonomy of spatial mitigation techniques. We analyze how spatial diversity interacts with envelope stability and CSI aging, and we evaluate the suitability of different techniques for passive sensing applications.

Third, we characterize \textbf{CSI aging} in cascaded backscatter channels using the Clarke-Jakes model and demonstrate that backscatter links decorrelate faster than conventional single-hop links. We analyze the coupling between CSI aging and both envelope stability and near-far mitigation, and we survey coherent and non-coherent detection methods under time-varying conditions.

This integrated treatment of waveform, spatial, and temporal dimensions under a unified sensing fidelity framework (BCF, SFI, SCR) is absent from the existing literature. By adopting a tag-centric perspective, we bridge the gap between communication-centric backscatter surveys and hardware-centric WPT surveys, providing a holistic view of sensing-aware backscatter system design.

The comparison in Table~\ref{tab:comparison} visually summarizes the coverage of existing surveys and highlights the unique contributions of this work.

\section{The Envelope Stability Challenge}
\label{sec:envelope}

Maintaining a stable RF envelope is a first-order design constraint for passive sensing. Unlike active receivers, backscatter tags lack automatic gain control (AGC) and wide dynamic range front-ends, rendering them highly sensitive to the large instantaneous power variations characteristic of multi-carrier signals. This section formalizes the envelope stability challenge, quantifies its impact on sensing fidelity, and surveys waveform design techniques through a tag-centric lens.

\subsection{Motivation and Problem Overview}
\label{sec:envelope_motivation}

Passive backscatter tags rely entirely on the incident RF waveform for energy harvesting, modulation, and sensing. Modern ambient illuminators, particularly OFDM-based systems, exhibit high peak-to-average power ratio (PAPR), resulting in deep envelope nulls and sharp power peaks. These fluctuations lead to three distinct degradation mechanisms:

\begin{itemize}
    \item \textbf{Energy starvation} during envelope nulls, where incident power falls below the rectifier activation threshold;
    \item \textbf{Rectifier saturation} during envelope peaks, generating harmonic and intermodulation distortion; and
    \item \textbf{Sensing fidelity degradation} due to envelope ripple, which biases resonant sensing elements and corrupts phase coherence.
\end{itemize}

While high-PAPR multicarrier waveforms can enhance RF-to-DC conversion efficiency at low input power levels by exploiting rectifier nonlinearity, the same peaks that boost harvested energy may also drive the rectifier into saturation under stronger illumination. This creates a fundamental tradeoff: waveforms optimized for wireless power transfer (WPT), such as multisine signals, may increase harvested DC power but simultaneously degrade sensing fidelity due to envelope instability. This tension is central to the \textbf{``Illuminator's Dilemma'':} the waveform parameters that maximize energy harvesting may be misaligned with those required for reliable sensing.

While ambient backscatter communication has been extensively surveyed \cite{van2022survey, memon2022survey, khan2022backscatter}, envelope stability is rarely treated as a first-order design objective. Existing surveys largely model the incident signal as a quasi-stationary excitation source, overlooking the impact of temporal envelope dynamics on tag operation and sensing accuracy \cite{wu2022survey}. This section explicitly elevates envelope stability to a central design dimension in sensing-aware backscatter networks.

\subsection{Impact of High-PAPR Waveforms on Tag Operation}
\label{sec:envelope_impact}


In OFDM systems, time-domain superposition of independently modulated subcarriers produces large envelope fluctuations \cite{jiang2009papr, hanzo2011ofdm}. The incident signal at the tag is given by:

\begin{equation}
x(t) = \Re\left\{ \sum_{n=0}^{N-1} X_n e^{j2\pi f_n t} \right\}
\label{eq:ofdm_envelope}
\end{equation}

where $X_n$ is the complex symbol on the $n$-th subcarrier. The corresponding passband waveform is $x_{\mathrm{RF}}(t) = \Re\{s(t)e^{j2\pi f_c t}\}$. The instantaneous envelope $r(t) = |s(t)|$ exhibits three regimes of interest relative to the tag's rectifier characteristics \cite{valenta2022rectenna, hemour2022nonlinear}:

\begin{itemize}
    \item \textbf{Starvation regime} ($r(t) < r_{\mathrm{th}}$): Insufficient forward bias; tag energy harvesting ceases and sensing echoes are missed \cite{collado2012waveform, valenta2014harvesting}.
    
    \item \textbf{Linear regime} ($r_{\mathrm{th}} \leq r(t) \leq r_{\mathrm{sat}}$): Normal operation; sensing accuracy is governed by envelope ripple \cite{boaventura2022waveform}.
    
    \item \textbf{Saturation regime} ($r(t) > r_{\mathrm{sat}}$): Deep conduction; harmonic mixing products corrupt sensing signatures \cite{ma2021waveform, kim2014wireless}.
\end{itemize}

For sensing-enabled tags, these effects manifest as resonance drift, missed echoes, and nonlinear mixing products, directly impairing sensing fidelity \cite{reindl2021saw, plessky2021saw}.

\subsection{The Illuminator's Dilemma: Active vs Passive Constraints}
\label{sec:illuminator_dilemma}

Active transmitters operate under spectral masks and bit error rate (BER) constraints but enjoy significant waveform freedom. Passive backscatter tags, in contrast, are governed by rectifier conduction thresholds, nonlinear I-V characteristics, and limited energy storage \cite{nikitin2010rfid, finkenzeller2010rfid}. This asymmetry, summarized in Table~\ref{tab:active_vs_passive}, explains the dominance of simple amplitude-based modulation schemes such as ASK and OOK in BLE Ambient IoT and related passive standards, where waveform simplicity reduces envelope sensitivity at the tag \cite{tao2018backscatter, dobbins2021energy}.

\begin{table}[H]
\centering
\caption{Comparison of Active Transmitter vs Passive Tag Constraints}
\label{tab:active_vs_passive}
\footnotesize
\renewcommand{\arraystretch}{1.2}
\setlength{\tabcolsep}{3pt}
\begin{tabular}{|p{2.0cm}|p{2.8cm}|p{2.8cm}|}
\hline
\textbf{Aspect} & \textbf{Active Transmitter} & \textbf{Passive Tag} \\
\hline
Dynamic range & Wide (AGC available) \cite{nikitin2010rfid} & Narrow (no AGC) \cite{finkenzeller2010rfid} \\
\hline
Power source & Onboard battery \cite{dobbins2021energy} & Harvested from incident RF \cite{valenta2022rectenna} \\
\hline
Waveform flexibility & High \cite{hanzo2011ofdm} & Constrained by rectifier \cite{nikitin2010rfid} \\
\hline
Sensitivity to envelope & Low & High (peaks and nulls) \cite{tao2018backscatter} \\
\hline
Common modulations & QAM, OFDM, OTFS \cite{hanzo2011ofdm, li2024isac} & ASK, OOK, PSK \cite{tao2018backscatter, dobbins2021energy} \\
\hline
\end{tabular}
\end{table}

High-PAPR multicarrier illumination thus creates a mismatch: the transmitter optimizes spectral efficiency, while the tag requires amplitude stability. Bridging this mismatch requires extending classical PAPR reduction into envelope-aware waveform engineering.

\subsection{Quantifying Envelope Stability: Tag-Centric Metrics}
\label{sec:bcf_metrics}

While PAPR is widely used to characterize waveform behavior at the transmitter, passive backscatter tags are primarily affected by the instantaneous envelope they receive. Consequently, sensing-aware backscatter systems benefit from metrics that explicitly capture envelope stability at the tag. Building on the metrics introduced in Section~\ref{sec:metrics}, we define the Backscatter Crest Factor (BCF) as a normalized measure of envelope spread at the tag location.

\subsubsection{Backscatter Crest Factor (BCF)}

The BCF is defined as the normalized peak-to-trough spread of the instantaneous envelope power:

\begin{equation}
\mathrm{BCF} = \frac{\displaystyle\max_{t} |s(t)|^2 - \min_{t} |s(t)|^2}{\mathbb{E}[|s(t)|^2]}
\label{eq:bcf}
\end{equation}

where $s(t)$ is the equivalent complex baseband signal of the incident waveform. This metric captures total envelope spread in the power domain. A low PAPR does not necessarily imply a low BCF, since reducing the peak amplitude alone does not eliminate deep envelope nulls. Consequently, two waveforms with similar PAPR may exhibit significantly different BCF values depending on their minimum envelope power \cite{gulia2025bctas}.

\subsubsection{Envelope Null Probability}

Let $r_{\mathrm{th}}$ denote the rectifier conduction threshold, where $r(t) = |s(t)|$ is the instantaneous envelope. The starvation probability is:

\begin{equation}
P_{\mathrm{null}} = \Pr\!\bigl(r(t) < r_{\mathrm{th}}\bigr)
\label{eq:null_prob}
\end{equation}

For large-subcarrier OFDM systems, where the envelope approximates a Rayleigh distribution with parameter $\sigma$:

\begin{equation}
P_{\mathrm{null}} \approx 1 - \exp\!\left(-\frac{r_{\mathrm{th}}^2}{2\sigma^2}\right)
\label{eq:null_rayleigh}
\end{equation}

The Rayleigh approximation is valid for conventional OFDM waveforms with a sufficiently large number of independently modulated subcarriers, where the central limit theorem yields approximately Gaussian in-phase and quadrature components. This quantifies the fraction of time the tag experiences insufficient excitation for reliable energy harvesting or sensing.

\subsubsection{Saturation Probability}

Let $r_{\mathrm{sat}}$ denote the rectifier nonlinear saturation threshold. The saturation probability is:

\begin{equation}
P_{\mathrm{sat}} = \Pr\!\bigl(r(t) > r_{\mathrm{sat}}\bigr)
\label{eq:sat_prob}
\end{equation}

Exceedances of $r_{\mathrm{sat}}$ drive the rectifier into deep conduction, generating harmonic and intermodulation products that contaminate the measurement spectrum \cite{boaventura2014boosting, ma2021waveform}.

\subsubsection{Relationship to PAPR, ESF, and BCF}

The relationships among PAPR, ESF, PMR, and BCF can be expressed as:

\begin{equation}
\begin{aligned}
\mathrm{PAPR} &= \frac{\max |s(t)|^2}{\mathbb{E}[|s(t)|^2]}, \\
\mathrm{PMR} &= \frac{\max |s(t)|^2}{\min |s(t)|^2}, \\
\mathrm{ESF} &= \frac{1}{\mathrm{PMR}} = \frac{\min |s(t)|^2}{\max |s(t)|^2}, \\
\mathrm{BCF} &= \frac{\max |s(t)|^2 - \min |s(t)|^2}{\mathbb{E}[|s(t)|^2]}
\end{aligned}
\label{eq:metric_definitions}
\end{equation}

It follows directly that:

\begin{equation}
\mathrm{BCF} = \mathrm{PAPR} \cdot (1 - \mathrm{ESF})
\label{eq:bcf_papr_esf}
\end{equation}

BCF combines the information captured by PAPR and ESF into a single normalized measure. Equation~(\ref{eq:bcf_papr_esf}) shows that the three metrics characterize complementary aspects of the waveform envelope. PAPR quantifies the severity of peak excursions relative to the average power, ESF quantifies envelope stability through the ratio of minimum-to-maximum power (or equivalently the reciprocal of PMR), and BCF captures the overall normalized envelope spread by jointly accounting for both peaks and troughs. While PAPR addresses peak excursions and ESF captures peak-to-floor stability, BCF provides a complete characterization of total envelope spread in the power domain, making it a useful figure of merit for evaluating waveform envelope stability in sensing-aware backscatter systems.

\subsection{PAPR Reduction Techniques: A Sensing-Aware Re-Evaluation}
\label{sec:papr_techniques}

The classical PAPR reduction literature treats PAPR as a transmitter-efficiency and spectral-containment problem, emphasizing mitigation of in-band distortion and out-of-band emissions under nonlinear amplification \cite{jiang2009papr, rahmatallah2013peak}. However, when high-PAPR multicarrier waveforms illuminate passive backscatter tags, the problem becomes fundamentally bidirectional. This subsection re-evaluates classical techniques through the lens of envelope stability.

\subsubsection{Distortion-Based Methods: Clipping and Companding}

Time-domain clipping is the simplest PAPR mitigation strategy. Comparative analyses \cite{humse2024papr, alrakah2024vlc} demonstrate measurable peak suppression but reaffirm the intrinsic trade-off between distortion and complexity. Clipping reduces saturation probability at both transmitter and tag but does not guarantee suppression of envelope nulls. Moreover, clipping noise may interfere with harmonic backscatter or resonant sensing channels.

\subsubsection{Probabilistic Methods: SLM and PTS}

Selected Mapping (SLM) and Partial Transmit Sequence (PTS) reduce PAPR by generating candidate waveforms through phase rotations and selecting the minimum-peak realization \cite{amhaimar2024slm, sinha2022slm, kumar2025papr}. These methods are distortionless but computationally intensive and require side-information overhead. Importantly, SLM and PTS minimize peak amplitude but do not explicitly bound envelope ripple or null probability—saturation events become less likely, but transient starvation intervals may remain unaddressed.

\subsubsection{Structured Methods: Tone Reservation and Active Constellation Extension}

Tone Reservation (TR) allocates specific subcarriers to generate time-domain cancellation signals, achieving distortionless peak suppression at the cost of spectral efficiency \cite{elhassan2020tr}. The significance of TR lies in its deterministic envelope control, enabling not only peak suppression but potentially variance control and null mitigation through appropriate optimization. Hybrid extensions such as TR-Index Modulation partially recover rate while maintaining PAPR gains \cite{nguyen2025trim}.

Active Constellation Extension (ACE) reduces PAPR by expanding constellation points within allowable decision regions rather than truncating amplitudes \cite{liu2019ace, wu2025asedmt}. ACE offers distortionless operation without spectral sacrifice. For sensing systems where spectral purity and harmonic integrity are critical, this distortionless characteristic is advantageous.

\subsubsection{Learning-Based Methods}

Recent work integrates machine learning into PAPR reduction. Neural-network approaches trained on optimization-generated data approximate low-PAPR solutions with reduced complexity \cite{dasilva2025nn, sharma2025aiml}. Unlike classical methods that minimize peak amplitude alone, neural approaches can incorporate arbitrary loss functions, opening the possibility of learning envelope-flatness objectives that penalize null probability and variance.

\subsubsection{ISAC-Inspired Optimization}

A further evolution appears in Integrated Sensing and Communication (ISAC) waveform optimization. In \cite{li2025ofdmisac}, PAPR is incorporated as a hard constraint alongside sensing-related waveform requirements within an ADMM-based optimization framework. Adjustable-PAPR design in \cite{bazzi2025isac} treats PAPR as a tunable parameter within a multi-objective sensing-communication optimization problem. These formulations embed envelope constraints directly into waveform design rather than applying PAPR reduction as post-processing.

\subsection{Waveform-Envelope-Sensing Suitability Mapping}
\label{sec:waveform_mapping}

Table~\ref{tab:waveform_sensing} maps major waveform classes to their envelope behavior and suitability for passive sensing deployments. Two key observations emerge: (i) no single waveform simultaneously controls peaks, nulls, and ripple—the three degradation channels identified above—and (ii) structured approaches (TR, ACE) and ISAC optimization show the most promise for extension toward comprehensive envelope stability engineering.

\begin{table*}[t]
\centering
\caption{Waveform-Envelope-Sensing Suitability Mapping}
\label{tab:waveform_sensing}
\footnotesize
\renewcommand{\arraystretch}{1.2}
\setlength{\tabcolsep}{4pt}
\begin{tabular}{|p{5.0cm}|p{3.9cm}|p{2.5cm}|p{3.8cm}|}
\hline
\textbf{Waveform / Technique} & \textbf{Envelope Behavior} & \textbf{Backscatter Suitability} & \textbf{Notes} \\
\hline
CW (Continuous Wave) \cite{clerckx2021fundamentals} & Flat ($\mathrm{BCF} \approx 0$) & Excellent & Ideal but spectrally inefficient \\
\hline
OFDM (baseline) \cite{jiang2009papr} & High peaks, deep nulls & Limited & Requires envelope shaping \\
\hline
OTFS \cite{li2024isac} & Moderate peaks & Moderate & Emerging; less studied \\
\hline
Multisine (WPT-optimized) \cite{clerckx2021fundamentals} & Intentionally designed with high peaks to exploit rectifier nonlinearity & Good for harvesting; may degrade sensing fidelity due to saturation & Optimized for DC; tradeoff with envelope stability \\
\hline
MD-OFDM \cite{gulia2025bctas} & Programmable subcarriers & Good & Enables BCF-aware selection \\
\hline
OFDM + Clipping \cite{humse2024papr} & Peak suppressed; distortion & Moderate & May introduce sensing artifacts \\
\hline
OFDM + SLM/PTS \cite{sinha2022slm, kumar2025papr} & Statistical peak reduction & Moderate & Nulls persist \\
\hline
OFDM + TR \cite{elhassan2020tr} & Deterministic peak control & Promising & Extensible to null/ripple control \\
\hline
OFDM + ACE \cite{liu2019ace} & Distortionless peak reduction & Good & Preserves spectral integrity \\
\hline
ISAC-optimized \cite{bazzi2025isac} & Tunable PAPR constraint & High & Directly extensible to BCF \\
\hline
\end{tabular}
\end{table*}

\subsection{Hardware Non-Linearity: The PA-Rectifier Cascade}
\label{sec:hardware_nonlinearity}

\begin{figure*}[H]
\centering
\includegraphics[width=0.8\textwidth]{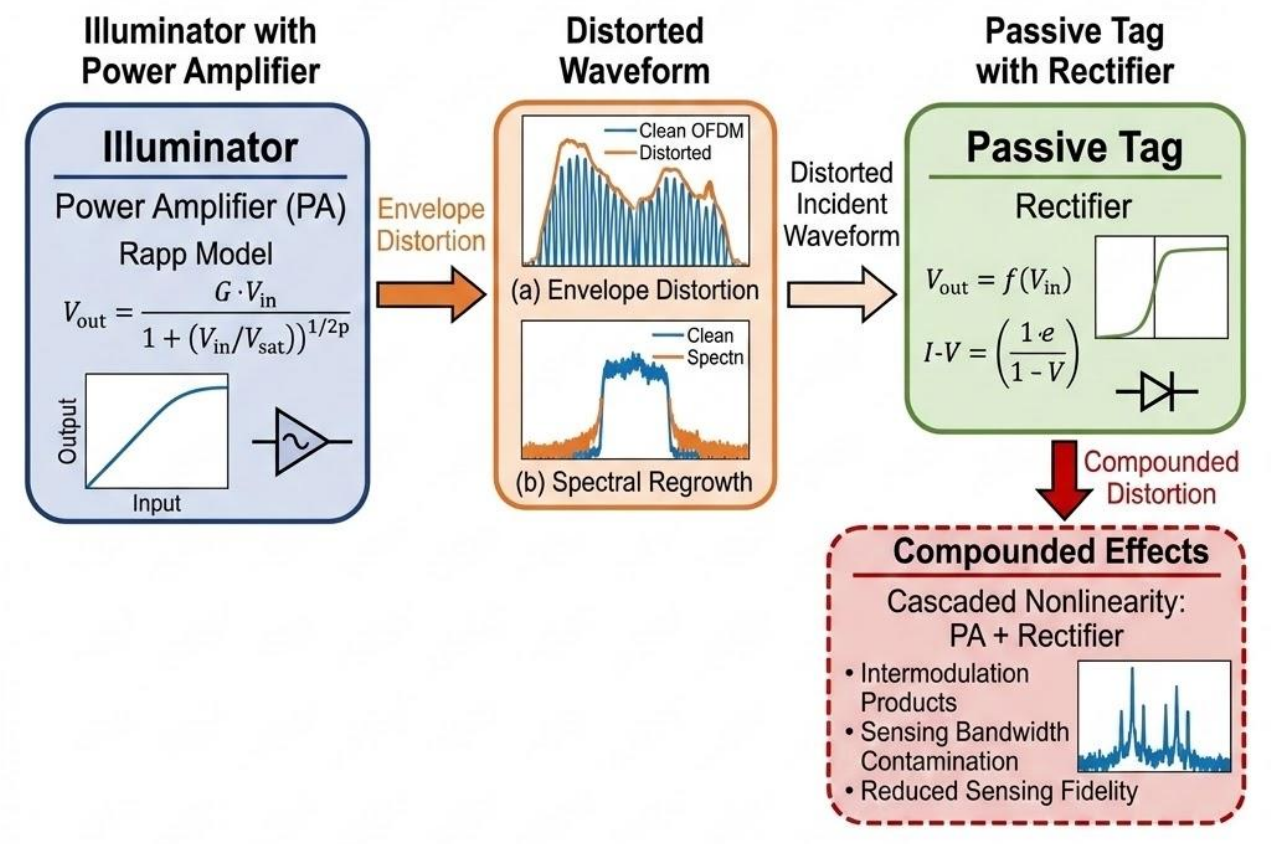}
\caption{Cascaded nonlinear effects: PA nonlinearity at the illuminator introduces envelope distortion and spectral regrowth, which is further shaped by the tag's rectifier nonlinearity, compounding envelope instability.}
\label{fig:pa_rectifier_cascade}
\end{figure*}

Non-ideal illuminator hardware further exacerbates envelope instability. Power amplifier (PA) nonlinearities introduce spectral regrowth and envelope distortion, which interact with rectifier nonlinearities at the tag, Figure~\ref{fig:pa_rectifier_cascade}. This compounded effect amplifies envelope instability and is largely ignored in idealized backscatter models \cite{rapp1991model, saleh_nonlinear}.

\subsubsection{Power Amplifier Non-Linearity: Rapp Model}

The Rapp model characterizes solid-state PA nonlinearity:

\begin{equation}
V_{\mathrm{out}}(t) = \frac{G V_{\mathrm{in}}(t)}{\left(1 + \left(\frac{|V_{\mathrm{in}}(t)|}{V_{\mathrm{sat}}}\right)^{2p}\right)^{1/(2p)}}
\label{eq:rapp}
\end{equation}

where $G$ is the linear gain, $V_{\mathrm{sat}}$ is the saturation voltage, and $p$ controls smoothness of the transition. PA nonlinearity produces spectral regrowth (out-of-band emissions) and compresses envelope peaks, altering the incident waveform before it reaches the tag.

\subsubsection{Rectifier Non-Linearity}

Recall from Section \ref{sec:metrics} the rectifier's nonlinear response:

\begin{equation}
V_{\mathrm{out}} = f(V_{\mathrm{in}}) \approx \sum_{k=0}^{\infty} \beta_k V_{\mathrm{in}}^k
\label{eq:rectifier_nonlinear}
\end{equation}

The cascaded PA-rectifier nonlinearity creates a compound effect where envelope distortion from the PA is further shaped by the tag's rectifier. This interaction is particularly problematic for sensing, as intermodulation products can fall within the sensing bandwidth \cite{hemour2022nonlinear, clerckx2023waveform}.

\subsection{Comparative Analysis and Open Challenges}
\label{sec:envelope_challenges}

Table~\ref{tab:envelope_summary} summarizes the techniques surveyed and their effectiveness across the three degradation dimensions.

\begin{table*}[t]
\centering
\caption{Summary of Envelope Stability Techniques}
\label{tab:envelope_summary}
\footnotesize
\renewcommand{\arraystretch}{1.2}
\setlength{\tabcolsep}{4pt}
\begin{tabular}{|p{8.8cm}|c|c|c|c|}
\hline
\textbf{Technique} & \textbf{Peak} & \textbf{Null} & \textbf{Ripple} & \textbf{Sensing} \\
\hline
Clipping \cite{humse2024papr, alrakah2024vlc} & Moderate & Poor & Poor & Low \\
\hline
SLM/PTS \cite{sinha2022slm, kumar2025papr} & Good & Poor & Poor & Moderate \\
\hline
Tone Reservation (TR) \cite{elhassan2020tr, nguyen2025trim} & Good & Fair & Fair & Good \\
\hline
ACE \cite{liu2019ace, wu2025asedmt} & Good & Poor & Poor & Moderate \\
\hline
ML-based \cite{dasilva2025nn, sharma2025aiml} & Good & Fair & Fair & High (potential) \\
\hline
ISAC optimization \cite{li2025ofdmisac, bazzi2025isac} & Excellent & Fair & Fair & High \\
\hline
MD-OFDM + BCF-TAS \cite{gulia2025bctas} & Excellent & Good & Good & High \\
\hline
\end{tabular}
\end{table*}

Key open challenges remain:

\begin{itemize}
    \item \textbf{Joint peak-null-ripple control}: No existing technique simultaneously addresses all three degradation mechanisms.
    \item \textbf{Hardware-aware optimization}: PA and rectifier nonlinearities are typically modeled separately; joint optimization is needed.
    \item \textbf{Mobility and channel dynamics}: Envelope stability at the tag depends on channel conditions, which change over time (bridging to Section~\ref{sec:csi_aging}).
    \item \textbf{Standards compatibility}: Ambient IoT standards (802.11bp, BLE) impose constraints on illuminator behavior.
\end{itemize}

These challenges motivate the spatial mitigation techniques surveyed in Section~\ref{sec:nearfar} and the temporal tracking methods in Section~\ref{sec:csi_aging}.

\subsection{Waveform Design Constraints in Backscatter Architectures}
\label{sec:waveform_constraints}

Waveform design in backscatter architectures is governed by a fundamental asymmetry between the active illuminator and the passive or semi-passive tag. In conventional active wireless systems, the transmitter generates its own RF carrier, controls the power amplifier, performs baseband processing, and can adapt its waveform, coding, modulation, beamforming, and resource allocation based on link conditions \cite{hanzo2011ofdm, jiang2009papr}. In contrast, a backscatter device does not generate an independent RF waveform. It communicates by modulating and reflecting an externally supplied carrier wave or ambient RF signal through changes in its antenna load impedance or backscatter coefficient \cite{yin2025, jantti2025, janjua2024, xu2024}. This architecture drastically reduces power consumption and hardware cost, but it also makes the tag highly dependent on the quality, stability, and structure of the incident waveform—a dependency that directly impacts the Envelope Stability Factor (ESF) and Backscatter Crest Factor (BCF) introduced in Section~\ref{sec:metrics}.

\subsubsection{Ambient IoT Device Classification and Constraints}

This distinction is particularly important in Ambient IoT. Recent 3GPP studies classify Ambient IoT devices according to energy storage capability, communication mechanism, design complexity, and power consumption \cite{yin2025, moloudi2025, abtouche2026, singh2025}. In this classification, the lowest-power devices rely on backscatter communication rather than active RF transmission. Device 1 and Device 2a are designed around backscatter operation, while Device 2b uses active RF transmission with higher power consumption \cite{yin2025, moloudi2025, abtouche2026, kim2025}. This means that passive and semi-passive Ambient IoT devices cannot be treated as reduced versions of conventional cellular modems. They operate with limited harvested energy, weak or intermittent energy availability, simplified RF front-ends, limited synchronization capability, and low computational resources \cite{singh2025, moloudi2025, kim2025, xin2024}.

Consequently, the waveform must be designed around the limitations of the tag rather than only around the spectral-efficiency objectives of the active network. This constraint directly motivates the tag-centric metrics introduced in Section~\ref{sec:metrics}—particularly the Envelope Stability Factor (ESF), which quantifies waveform flatness, and the Sensing Fidelity Index (SFI), which captures sensing degradation under unstable illumination.

\subsubsection{Tag-Side Modulation and Implementation Simplicity}

The same constraint explains why simple tag-side modulation remains dominant in backscatter systems. In RFID-inspired and Ambient IoT designs, amplitude shift keying (ASK), on-off keying (OOK), and related line-coding schemes are widely used because they can be implemented using low-complexity impedance switching and simple envelope detection \cite{yin2025, singh2025, dobbins2021energy, nikitin2010rfid}. In OOK, the tag represents information through the presence or absence of a reflected component. In ASK backscatter, information bits are mapped through different amplitudes of the backscatter coefficient, while PSK uses phase variations of the reflected signal \cite{yin2025, finkenzeller2010rfid}.

\begin{figure*}[H]
\centering
\includegraphics[width=0.8\textwidth]{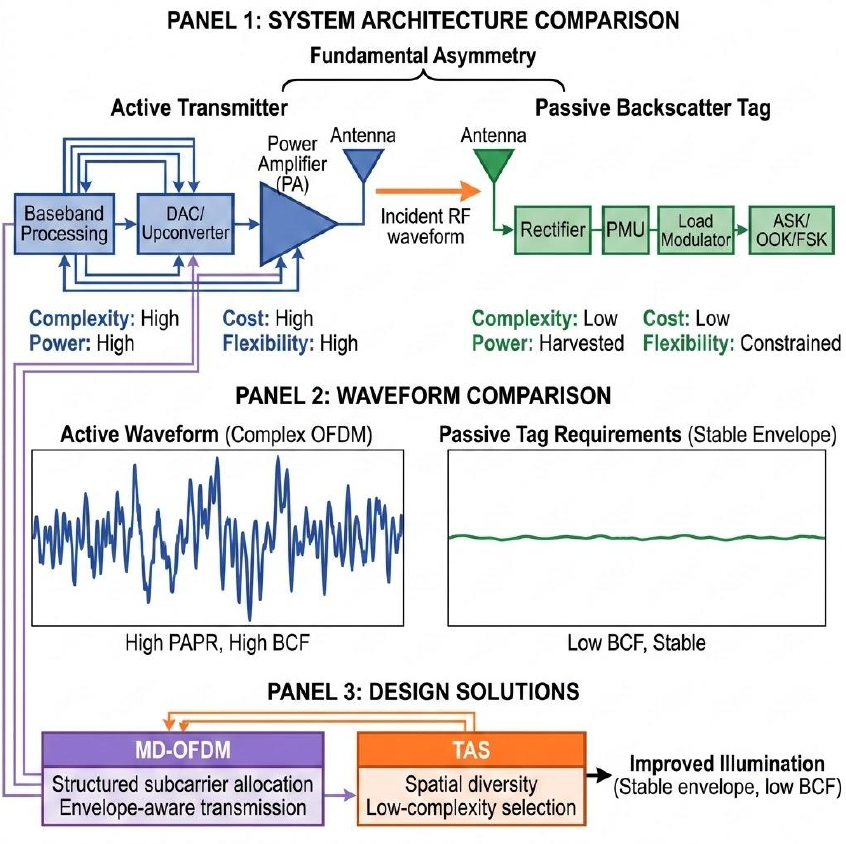}
\caption{Waveform design constraints in passive backscatter architectures. Unlike conventional active radios, passive tags rely on externally illuminated waveforms and therefore require simple modulation schemes, stable incident envelopes, and low-complexity hardware. The figure highlights the asymmetry between the active transmitter and passive tag and illustrates how transmitter-side techniques such as multidimensional OFDM (MD-OFDM) and transmit antenna selection (TAS) can improve illumination quality while preserving the simplicity of the backscatter device.}
\label{fig:waveform_constraints}
\end{figure*}

Figure~\ref{fig:waveform_constraints} illustrates the fundamental asymmetry between active transmitters and passive tags, highlighting why simple modulation schemes and stable incident envelopes are essential for passive backscatter devices.

Although PSK may offer energy-efficiency advantages, it requires more difficult impedance matching and therefore more complex hardware implementation \cite{yin2025, nikitin2010rfid}. RFID line codes such as PIE, FM0, and Miller-modulated subcarrier further show how practical tag-side waveforms prioritize self-synchronization and implementation simplicity rather than high-order modulation or strong error-correction capability \cite{yin2025, dobbins2021energy, griffin2022multistatic}.

\subsubsection{Fragility of Amplitude-Based Modulation}

However, simple amplitude-based modulation also exposes the fragility of passive tags. Since OOK and ASK depend strongly on reflected amplitude, their reliability can degrade under fading, interference, envelope variation, and direct-link interference from the primary signal \cite{tao2018backscatter, ma2021waveform, boaventura2022waveform}. This is particularly problematic for sensing-aware tags, where the Sensing-to-Clutter Ratio (SCR) depends on the stability of the reflected amplitude relative to direct-path interference (see Section~\ref{sec:metrics}).

This is why recent works have investigated frequency-domain and delay-domain approaches such as FSK and DSK. In OFDM-based symbiotic radio, Janjua et al. proposed the use of FSK-based backscatter modulation with deliberately emptied OFDM subcarriers, allowing the shifted backscatter signal and the primary OFDM signal to remain orthogonal \cite{janjua2024}. Their design also includes an OOK-based scheme in which the backscatter device shifts the primary data to null subcarriers when transmitting bit ``1'' and remains constant for bit ``0'' \cite{janjua2024, elmossallamy2018, elmossallamy2019}.

Similarly, Uledi et al. used delay shift keying (DSK) within a zero-tail DFT-s-OFDM waveform, allowing backscatter information to be embedded in deterministic delay shifts inside the zero-tail region \cite{uledi2026}. These approaches show that reliability can be improved by designing waveform resources that separate the weak backscatter signal from the stronger primary signal, while still keeping the tag-side operation lightweight.

\subsubsection{OFDM and Symbiotic Radio: Opportunities and Challenges}

The modulation choice is therefore inseparable from the incident waveform. OFDM is attractive for backscatter because it is already used in cellular and Wi-Fi systems, making it suitable for symbiotic and ambient operation \cite{janjua2024, uledi2026, abtouche2026, hanzo2011ofdm}. By reusing existing infrastructure, backscatter devices avoid the need for dedicated carrier generation, reducing deployment and hardware cost \cite{jantti2025, kim2025, xin2024, liu2022ambient}.

\begin{figure}
\centering
\hspace*{-10mm}
\includegraphics[width=0.50\textwidth]{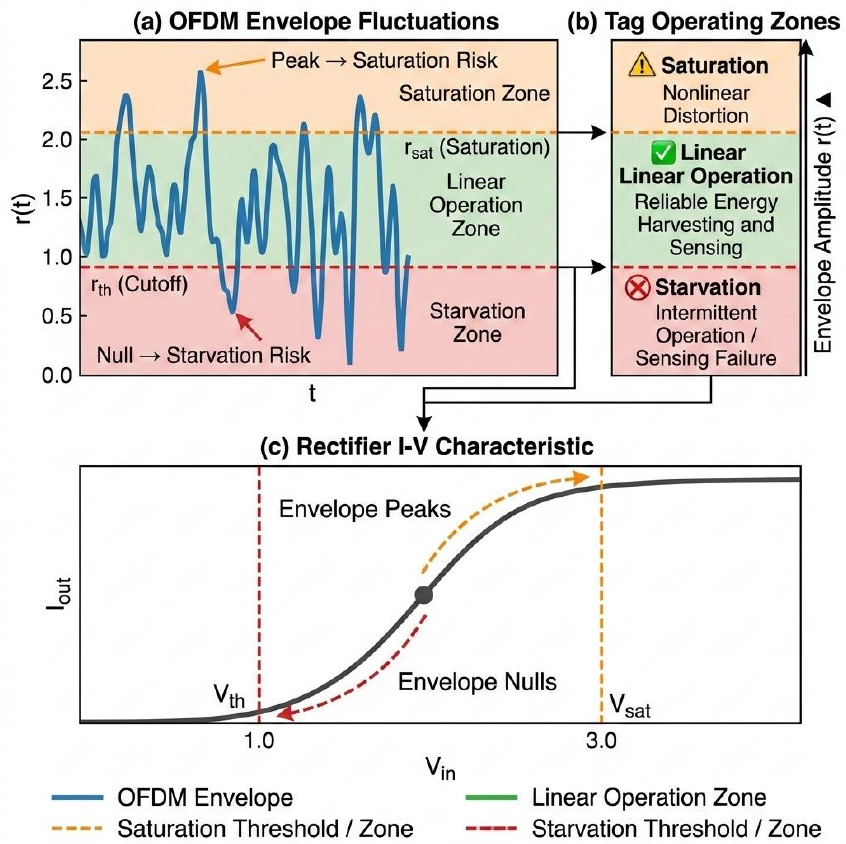}
\caption{Impact of envelope fluctuations on passive backscatter tags. High-PAPR waveforms produce large envelope fluctuations that may lead to rectifier saturation under excessive incident power or energy starvation under deep envelope nulls, whereas stable envelope regions enable reliable energy harvesting, sensing, and communication.}
\label{fig:envelope_stability}
\end{figure}

Nevertheless, OFDM also introduces important constraints. Its multicarrier structure produces time-domain envelope fluctuations and high peak-to-average power ratio (PAPR), which can be problematic for passive devices that depend on nonlinear rectification, impedance switching, and weak reflected signals \cite{jiang2009papr, rahmatallah2013peak, collado2012waveform}. Deep envelope nulls may reduce harvested energy or weaken the backscatter response, while large peaks may drive the tag front-end into nonlinear operation—directly impacting the Envelope Stability Factor (ESF) and Backscatter Crest Factor (BCF) introduced earlier, as shown in Figure~\ref{fig:envelope_stability}.

Thus, for backscatter architectures, waveform design cannot be evaluated only by transmitter-side metrics such as spectral efficiency, BER, or PAPR. It must also consider the tag-side quality of excitation, including envelope stability, detectability of the reflected signal, and compatibility with the rectifier and impedance-switching network. This tag-side perspective is the central theme of this survey.

\subsubsection{Emerging Symbiotic Radio Designs}

Recent symbiotic radio designs address this problem by moving complexity away from the tag and into the active network. In \cite{janjua2024}, the base station reshapes the OFDM resource grid by reserving null subcarriers so that the backscatter signal can be shifted into interference-free spectral locations. In \cite{uledi2026}, the zero-tail portion of ZT DFT-s-OFDM is exploited as a protected time-domain resource for backscatter transmission, avoiding direct-link interference while preserving compatibility with conventional receivers.

In \cite{abtouche2026}, an OFDM-AFDM symbiotic framework is proposed for integrated sensing and backscatter communication, where chirp-based AFDM reflections exploit waveform-domain orthogonality, affine-domain sparsity, and processing gain to improve weak backscatter detection and suppress inter-backscatter-device interference. Across these works, the same design principle emerges: the active transmitter or reader should shape the waveform environment so that the passive tag can remain simple.

This principle motivates multidimensional OFDM combined with transmit antenna selection (TAS). MD-OFDM provides additional waveform-domain degrees of freedom by distributing information and energy across structured time-frequency dimensions. TAS adds spatial diversity by selecting the transmit antenna whose channel, coupling profile, or illumination condition is most favorable to the passive tag. Unlike full beamforming, TAS does not require continuous phase-coherent precoding or multiple active RF chains at the tag \cite{sanayei2007mimo, gorokhov2002antenna, chen2022antenna, vicario2022antenna}. The tag remains a low-complexity impedance-switching device, while the active transmitter performs the selection process. This is consistent with the broader Ambient IoT design philosophy, where the network, reader, gNB, or UE absorbs complexity so that passive devices can maintain low power and low cost \cite{yin2025, jantti2025, kim2025, xin2024}.

\subsubsection{Joint Waveform-Spatial Design: MD-OFDM with TAS}

The benefit of MD-OFDM with TAS is that it jointly addresses waveform and spatial fragility. From the waveform side, MD-OFDM can support structured subcarrier allocation, diversity, and envelope-aware transmission \cite{gulia2025bctas, ma2021waveform, li2024isac}. From the spatial side, TAS can select an antenna that reduces unfavorable fading, improves incident power balance, or enhances backscatter detectability. This is especially useful because passive tags cannot easily perform adaptive equalization, power control, or interference cancellation locally. Therefore, MD-OFDM and TAS shift the burden of adaptation to the active transmitter, where energy and processing resources are available.

In a sensing-aware backscatter system, this can help stabilize the RF excitation seen by the tag, reduce the risk of starvation or saturation, and improve the reliability of the weak reflected signal. This directly impacts the Sensing Fidelity Index (SFI) and Sensing-to-Clutter Ratio (SCR) defined in Section~\ref{sec:metrics}, as well as the Backscatter Crest Factor (BCF) introduced in Section~\ref{sec:bcf_metrics}.

\subsubsection{Standards Implications and Future Directions}

These constraints also have direct standardization implications. The reviewed 3GPP Ambient IoT works emphasize low cost, low power, low complexity, energy harvesting, waveform design, modulation, coding, random access, duty-cycled monitoring, coexistence with NR, and protocol simplification \cite{yin2025, singh2025, moloudi2025, kim2025, xin2024}. Cellular Ambient IoT further aims to reuse existing cellular infrastructure, gNBs, UEs, and intermediate nodes to support backscatter devices without requiring a fully independent network for passive tags \cite{jantti2025, xin2024, wang2024ambient}.

Although the reviewed papers focus mainly on 3GPP Ambient IoT, the same physical-layer constraints are relevant to IEEE 802.11bp and BLE Ambient IoT. In all of these ecosystems, passive or near-passive devices must communicate without the hardware freedom of active radios \cite{tao2018backscatter, dobbins2021energy, liu2022ambient}. Hence, waveform designs for such standards must preserve compatibility with existing infrastructure while supporting simple tag-side modulation, stable RF excitation, and reliable weak-signal detection.

\subsubsection{Toward Tag-Aware Illumination Engineering}
Consequently, future backscatter waveform design should move from transmitter-centric optimization toward tag-aware illumination engineering. Classical active-radio metrics such as throughput, BER, spectral containment, and PAPR remain important, but they are insufficient for passive sensing-aware architectures. A backscatter-oriented waveform must also be judged by whether it:

\begin{itemize}
    \item Supports low-complexity ASK/OOK/FSK/DSK implementation,
    \item Reduces direct-link and inter-device interference,
    \item Preserves sufficient harvested energy, and
    \item Maintains a stable incident envelope at the tag (quantified by ESF and BCF).
\end{itemize}

Recent works on low-complexity Ambient IoT waveform-modulation-coding design \cite{yin2025}, physical-layer standardization \cite{singh2025, moloudi2025}, cellular integration \cite{jantti2025, xin2024}, OFDM-based interference-free backscatter \cite{janjua2024}, ZT DFT-s-OFDM delay-domain backscatter \cite{uledi2026}, OFDM-AFDM sensing-aware coexistence \cite{abtouche2026}, and long-range 2FSK ambient FM backscatter \cite{xu2024} collectively support this direction. 

MD-OFDM with TAS fits naturally into this trend because it improves reliability through transmitter-side waveform and spatial selection while preserving the simplicity of passive backscatter tags. This joint waveform-spatial design approach directly addresses the three challenges of this survey: envelope stability (via BCF-aware waveform design), near-far mitigation (via TAS for spatial power balancing), and CSI aging (via TAS selection based on channel state).

\subsection{Summary}
\label{sec:envelope_summary}

This section has formalized the envelope stability challenge in sensing-aware backscatter communications. We introduced tag-centric metrics including the Backscatter Crest Factor (BCF), quantified the impact of envelope fluctuations on sensing fidelity, re-evaluated classical PAPR reduction techniques through a sensing-aware lens, and identified open challenges. The key insight is that envelope stability requires more than limiting peak amplitude—it requires controlling the full amplitude distribution at the tag location. This motivates the waveform-spatial-temporal co-design framework that organizes the remainder of this survey.

However, waveform design alone has fundamental limits. Even an optimally shaped envelope can be severely distorted by channel fading—a deep null at the tag's location renders any waveform ineffective. This vulnerability provides the natural motivation for spatial diversity techniques: if the channel acts as a filter, spatial selection techniques—such as transmit antenna selection (TAS), beamforming, or intelligent reflecting surfaces (IRS)—can choose the signal path that minimizes the Backscatter Crest Factor (BCF) at the tag, effectively 'flattening' the envelope through spatial diversity rather than waveform shaping alone. Section~\ref{sec:nearfar} surveys these spatial mitigation strategies.

\section{The Near-Far Interference Gap}
\label{sec:nearfar}

In dense industrial deployments such as automated warehouses and logistics hubs, passive backscatter tags experience a pronounced \emph{near-far effect} due to the uneven spatial distribution of incident RF power. Unlike active receivers, battery-free tags lack automatic gain control (AGC) and linear front-end amplification, making them highly sensitive to power disparities arising from distance, antenna radiation patterns, polarization mismatch, blockage, and multipath propagation \cite{liu2013ambient, kimionis2014bistatic, griffin2009multipath}. This section formalizes the near-far challenge, surveys spatial mitigation techniques, and provides a comparative analysis of existing approaches.

\subsection{Physical Origin: Saturation vs. Starvation}
\label{sec:nearfar_origin}

Passive tags operate within a narrow window dictated by the nonlinear current-voltage characteristics of RF-to-DC rectifier circuits \cite{nikitin2010rfid, clerckx2021fundamentals, valenta2022rectenna}. Two primary regimes emerge:

\begin{itemize}
    \item \textbf{Near-field saturation:} Tags located close to the illuminator experience excessive incident power, driving rectifiers into saturation and generating nonlinear distortion in the backscattered signal.
    \item \textbf{Far-field starvation:} Tags at coverage edges receive insufficient RF energy to exceed the rectifier turn-on threshold, resulting in intermittent operation or complete sensing failure \cite{nikitin2006path}.
\end{itemize}

Unlike active wireless systems, where near-far primarily affects signal-to-noise ratio (SNR) at the receiver, in passive sensing networks it manifests as a \emph{tag-side envelope dynamic-range limitation}, as shown in Figure~\ref{fig:envelope_stability}.


\subsubsection{Quantitative Near-Far Model}

To quantify the severity of the near-far effect, consider a tag at position $\mathbf{r}_k$. The incident power is:

\begin{equation}
P_k = P_t G_t(\mathbf{r}_k) L(\mathbf{r}_k) |h_k|^2
\label{eq:incident_power}
\end{equation}

where $P_t$ is the transmit power, $G_t(\mathbf{r}_k)$ is the antenna gain, $L(\mathbf{r}_k)$ models large-scale path loss, and $h_k$ captures small-scale fading.

Reliable sensing requires:

\begin{equation}
P_{\mathrm{th}} \le P_k \le P_{\mathrm{sat}}
\label{eq:dynamic_range}
\end{equation}

For $K$ tags, the fractions experiencing starvation or saturation are:

\begin{equation}
\eta_{\mathrm{starve}} = \frac{|\{k : P_k < P_{\mathrm{th}}\}|}{K}, \quad
\eta_{\mathrm{sat}} = \frac{|\{k : P_k > P_{\mathrm{sat}}\}|}{K}
\label{eq:fractions}
\end{equation}

In backscatter systems, the effective path loss is inherently two-way, as the signal propagates from illuminator to tag and then back to the reader:

\begin{equation}
P_k \propto P_t G_t G_r \, L_{IT}(\mathbf{r}_k) L_{TR}(\mathbf{r}_k) |\Gamma_k|^2
\label{eq:twoway_pathloss}
\end{equation}

where $L_{IT}(\mathbf{r}_k)$ and $L_{TR}(\mathbf{r}_k)$ are the illuminator-to-tag and tag-to-reader path losses, respectively, and $\Gamma_k$ is the tag's reflection coefficient. The product $L_{IT} L_{TR}$ captures the two-way nature of the backscatter link, which results in significantly stronger attenuation than conventional one-way wireless links. The near-far power ratio is:

\begin{equation}
\Gamma_{\mathrm{NF}} = \frac{\max_k P_k}{\min_k P_k}
\label{eq:nearfar_ratio}
\end{equation}

Large $\Gamma_{\mathrm{NF}}$ values indicate severe disparity, where some tags saturate while others starve.

\subsection{Spatial Mitigation Techniques: A Comparative Survey}
\label{sec:spatial_mitigation}

Near-far mitigation has been extensively studied across multiple layers. Existing approaches can be broadly classified into power-domain, spatial-domain, and protocol-domain techniques \cite{liu2022ambient, van2022survey}. Each addresses different aspects of RF power imbalance but has distinct limitations.

\subsubsection{Power-Domain Techniques}

Power-domain techniques regulate transmitted or incident RF energy to maintain tag operation within the rectifier's dynamic range \cite{clerckx2021fundamentals}.

\textbf{Transmit power control:} Adjusting global transmit power avoids saturation at nearby tags but reduces energy for distant tags, exacerbating starvation. This creates a fundamental tradeoff.

\textbf{Waveform shaping:} Reducing peak-to-average power ratio (PAPR) limits instantaneous envelope excursions, mitigating rectifier saturation \cite{ma2021waveform, collado2012waveform}. However, these techniques primarily address temporal fluctuations and do not provide spatial control over power distribution, limiting effectiveness in heterogeneous deployments.

\subsubsection{Multi-Antenna Beamforming}

Beamforming directs energy toward specific spatial directions using phase-coherent precoding \cite{Xiang2012RobustBeamformingWIPT, Zeng2015OptimizedTrainingWET}. The received power at tag $k$ becomes:

\begin{equation}
P_k = |\mathbf{h}_k^H \mathbf{w}|^2 P_t
\label{eq:beamforming}
\end{equation}

where $\mathbf{w}$ is the precoding vector and $\mathbf{h}_k$ is the channel vector. Classical strategies maximize total harvested energy:

\begin{equation}
\max_{\mathbf{w}} \sum_k P_k
\label{eq:beamforming_objective}
\end{equation}

However, this formulation does not constrain peak exposure, potentially increasing saturation at nearby tags. Beamforming also requires accurate CSI, introduces higher PAPR, and increases hardware complexity.

\subsubsection{Intelligent Reflecting Surfaces (IRS)}

IRS technology shapes the wireless environment by controlling reflected paths \cite{direnzo2019smart, guo2020irs}. The received power becomes:

\begin{equation}
P_k = \left| \mathbf{h}_{d,k} + \mathbf{h}_{r,k}^H \mathbf{\Phi} \mathbf{G} \right|^2 P_t
\label{eq:irs}
\end{equation}

where $\mathbf{\Phi}$ is a diagonal matrix of adjustable phase shifts. IRS optimization typically maximizes average coverage \cite{basar2020wireless, wu2021irs}:

\begin{equation}
\max_{\mathbf{\Phi}} \sum_k P_k
\label{eq:irs_objective}
\end{equation}

While IRS improves coverage and mitigates deep fading, achieving fine-grained balancing across heterogeneous tag sensitivities requires high-resolution control and frequent adaptation, increasing deployment complexity.

\subsubsection{Cooperative Tag Relaying}

Cooperative backscatter allows well-illuminated tags to assist energy-starved neighbors \cite{liu2013ambient, zamora2022cooperative}. If tag $A$ relays energy to tag $B$:

\begin{equation}
P_B = \beta_A P_A |h_{A,B}|^2
\label{eq:cooperative}
\end{equation}

where $\beta_A$ is the reflection coefficient of tag $A$. While cooperative relaying reduces starvation at coverage edges, nonlinear distortion at upstream saturated tags ($P_A > P_{\mathrm{sat}}$) may propagate through the relay path and degrade sensing fidelity. Multi-hop reflection chains also suffer from cumulative path loss.

\subsubsection{Multi-Reader and Multistatic Diversity}

Deploying multiple illuminators introduces macro-diversity \cite{griffin2022multistatic, arnitz2013multistatic}. The aggregate received power is:

\begin{equation}
P_k = \sum_{m=1}^{M} P_t^{(m)} G_m(\mathbf{r}_k) L_m(\mathbf{r}_k)
\label{eq:multireader}
\end{equation}

Spatial diversity reduces the probability of persistent starvation $\Pr(P_k < P_{\mathrm{th}})$ but may increase aggregate exposure beyond $P_{\mathrm{sat}}$ in near-field regions. Multi-reader architectures improve robustness against blockage but do not inherently enforce dynamic-range constraints and increase infrastructure cost.

\subsubsection{Transmit Antenna Selection (TAS)}

Transmit Antenna Selection (TAS) exploits spatial diversity without requiring phase-coherent precoding \cite{sanayei2007mimo, gorokhov2002antenna}. Each transmit antenna exhibits a distinct spatial coupling profile due to geometry, polarization, and multipath. 

\textit{Remark:} In flat-fading channels, antenna selection changes only the received signal magnitude through scalar channel gain and therefore does not alter normalized temporal envelope metrics such as BCF. The discussion in this survey concerns frequency-selective propagation, where different antenna selections may produce different effective channel responses and consequently different received envelope characteristics.

TAS selectively activates antennas to achieve:
\begin{itemize}
    \item \textbf{Spatial nulling:} Reducing incident power at near-field tags
    \item \textbf{Energy steering:} Enhancing illumination toward distant tags
\end{itemize}

\begin{figure*}[H]
\centering
\includegraphics[width=0.8\textwidth]{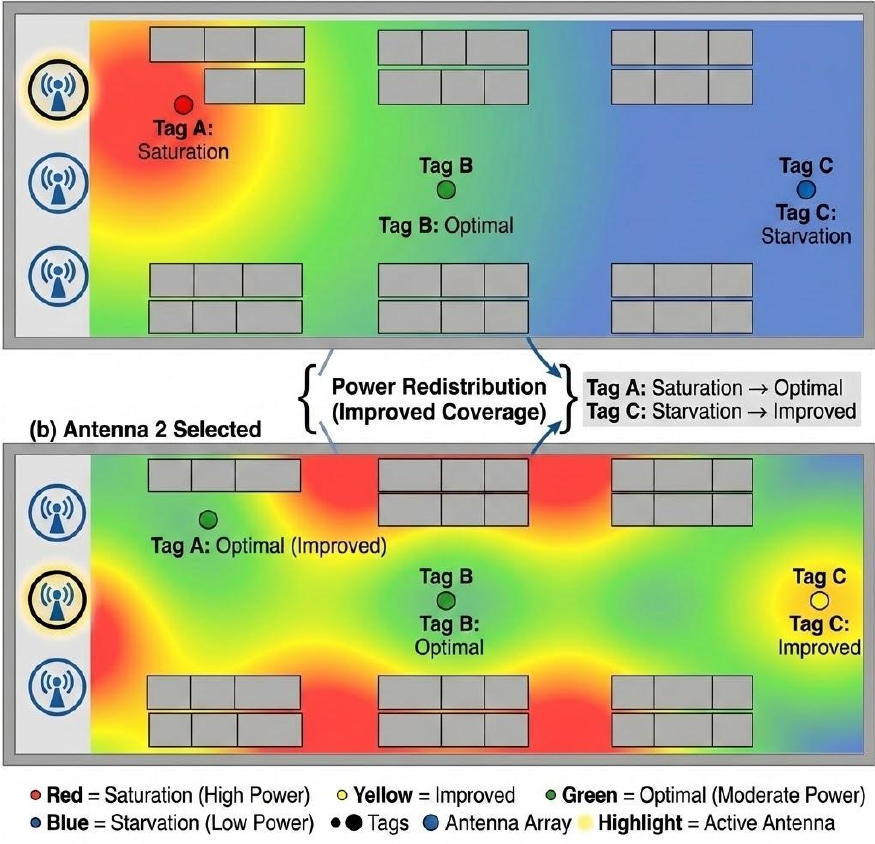}
\caption{Conceptual heatmap of RF illumination under different antenna selections. Under one antenna configuration, Tag A may experience saturation while Tag C remains energy starved. When another antenna is selected, the spatial power distribution changes, potentially balancing the incident RF envelope across tags.}
\label{fig:tas_heatmap}
\end{figure*}

Compared to beamforming, TAS requires no CSI for precoding (only for selection), preserves low-PAPR characteristics, and has lower hardware complexity. However, its effectiveness depends on antenna diversity, channel richness, and update frequency \cite{chen2022antenna, vicario2022antenna}. In dense deployments, limited spatial degrees of freedom may not balance all tags simultaneously, as shown in Figure~\ref{fig:tas_heatmap}.

\subsection{Comparative Analysis}
\label{sec:nearfar_comparison}

Table~\ref{tab:nearfar_comparison} compares spatial mitigation techniques across key dimensions.

\begin{table}[H]
\centering
\caption{Comparison of Near-Far Mitigation Techniques}
\label{tab:nearfar_comparison}
\footnotesize
\renewcommand{\arraystretch}{1.2}
\setlength{\tabcolsep}{3pt}
\begin{tabular}{|p{2.9cm}|p{1.2cm}|p{1.2cm}|p{1.8cm}|}
\hline
\textbf{Technique} & \textbf{CSI Req.} & \textbf{Infra. Cost} & \textbf{Sensing Suitability} \\
\hline
Power Control \cite{clerckx2021fundamentals} & Minimal & Low & Low \\
\hline
Beamforming \cite{Xiang2012RobustBeamformingWIPT} & High & Moderate-High & Moderate \\
\hline
IRS \cite{basar2020wireless, wu2021irs} & Moderate & High & Moderate \\
\hline
Cooperative Relaying \cite{liu2013ambient, zamora2022cooperative} & Moderate & Low (tags) & Low-Moderate \\
\hline
Multi-Reader \cite{griffin2022multistatic} & Low & High & Moderate \\
\hline
TAS \cite{sanayei2007mimo, chen2022antenna} & Low-Moderate & Low & Moderate-High \\
\hline
\end{tabular}
\end{table}

Table~\ref{tab:nearfar_summary} summarizes representative qualitative trends for starvation/saturation mitigation and sensing fidelity.

\begin{table}[H]
\centering
\caption{Qualitative Performance Summary of Near-Far Mitigation Techniques}
\label{tab:nearfar_summary}
\footnotesize
\renewcommand{\arraystretch}{1.2}
\setlength{\tabcolsep}{3pt}
\begin{tabular}{|p{3.2cm}|p{1.5cm}|p{1.5cm}|p{1.2cm}|}
\hline
\textbf{Technique} & \textbf{Starvation Mitigation} & \textbf{Saturation Mitigation} & \textbf{Sensing Fidelity} \\
\hline
Power Control \cite{clerckx2021fundamentals} & Poor & Moderate & Low \\
\hline
Beamforming \cite{Xiang2012RobustBeamformingWIPT} & Good & Poor-Moderate & Moderate \\
\hline
IRS \cite{basar2020wireless, wu2021irs} & Good & Moderate & Moderate \\
\hline
Cooperative Relaying \cite{liu2013ambient, zamora2022cooperative} & Moderate & Poor & Low-Moderate \\
\hline
Multi-Reader \cite{griffin2022multistatic} & Good & Poor-Moderate & Moderate \\
\hline
TAS \cite{sanayei2007mimo, chen2022antenna} & Moderate-Good & Moderate-Good & Moderate-High \\
\hline
\end{tabular}
\end{table}

\subsection{Open Challenges and Research Directions}
\label{sec:nearfar_challenges}

Several open challenges remain:

\begin{itemize}
    \item \textbf{Dynamic deployment and mobility:} In warehouse environments, pallet movement, robotic platforms, and human activity continuously reshape multipath conditions. Designing low-overhead spatial adaptation mechanisms without frequent CSI acquisition remains an open problem \cite{ma2022channel, kim2021channel}.
    
    \item \textbf{Heterogeneous rectifier sensitivity:} Passive tags exhibit heterogeneous thresholds due to manufacturing variation and sensing modality. A mitigation strategy optimized for one dynamic range may drive other tags into saturation or starvation.
    
    \item \textbf{Hardware nonlinearity coupling:} Power amplifier nonlinearity and rectifier nonlinearity interact in non-trivial ways in multi-carrier systems. Spatial mitigation may alter peak distributions and amplify spectral regrowth \cite{hemour2022nonlinear, rapp1991model}.
    
    \item \textbf{Cross-layer optimization:} Waveform shaping, antenna selection, and index modulation all influence envelope behavior across spatial and temporal domains. Joint optimization across these layers remains largely unexplored.
    
    \item \textbf{Interference and coexistence:} Industrial deployments coexist with WiFi, BLE, and other ambient RF sources. Understanding how spatial mitigation interacts with coexistence policies is essential for practical deployment.
    
    \item \textbf{Scalability in dense networks:} As tag density increases, the probability that some tags lie outside the acceptable dynamic range grows. Characterizing scaling laws for near-far mitigation remains an open theoretical question.
\end{itemize}

These challenges motivate the temporal tracking and non-coherent detection techniques surveyed in Section~\ref{sec:csi_aging}.

\subsection{Summary}
\label{sec:nearfar_summary}

This section has formalized the near-far interference gap in sensing-aware backscatter communications. We introduced quantitative metrics for spatial power disparity, surveyed classical and emerging mitigation techniques, and provided a comparative analysis of their strengths and limitations. No single technique fully resolves the simultaneous challenges of energy starvation and rectifier saturation in dense deployments. The choice of technique depends on deployment constraints, hardware capabilities, and sensing requirements. This motivates the waveform-spatial-temporal co-design framework that organizes the remainder of this survey.

\section{Channel State Information Aging in Passive Backscatter Networks}
\label{sec:csi_aging}

\begin{figure}
\centering
\includegraphics[width=0.50\textwidth]{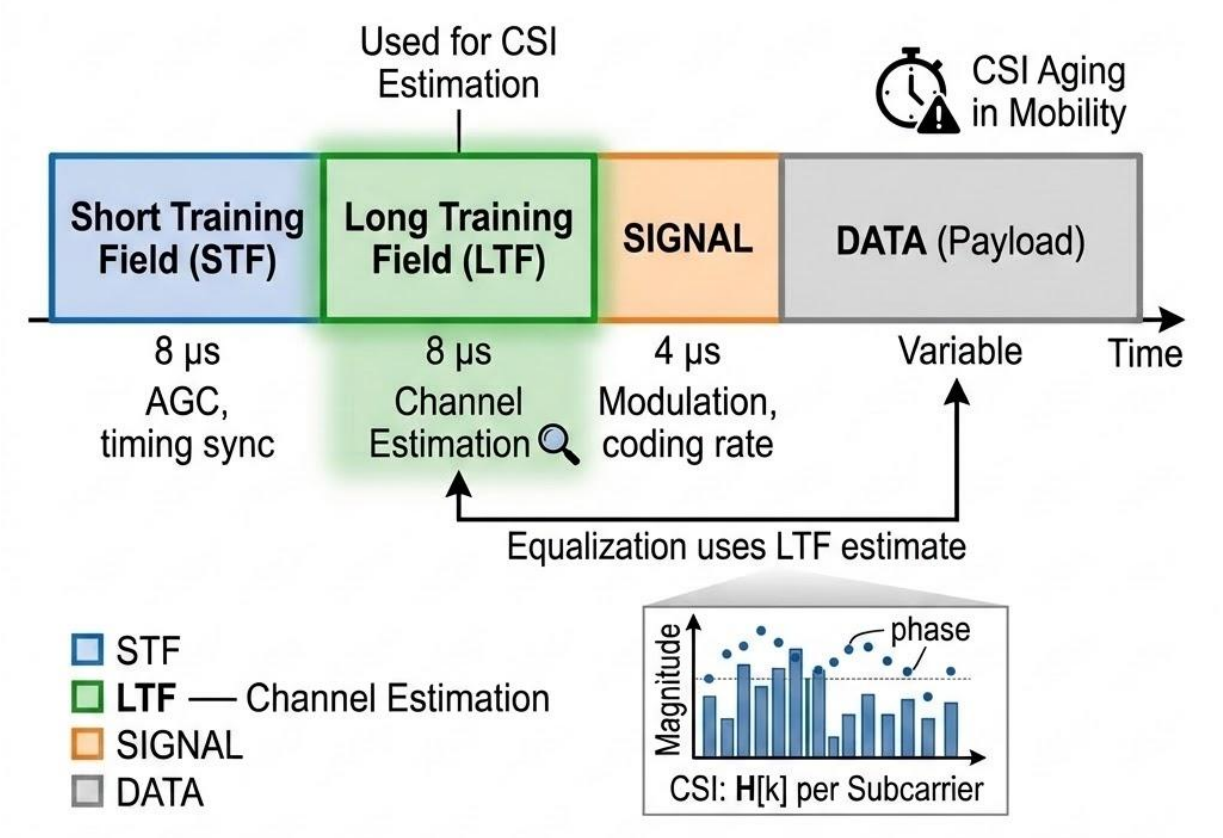}
\caption{802.11a PHY frame format illustrating pilot-based channel estimation using the Long Training Field (LTF). The estimated CSI is then used for equalization of subsequent data symbols within the frame.}
\label{fig:80211a_frame}
\end{figure}

Channel State Information (CSI) describes how a wireless signal changes as it travels from the transmitter to the receiver. As the signal moves through the air, it can get weaker, bounce off objects, arrive at slightly different times, or shift in phase. CSI captures these effects by showing how the channel affects the signal's strength (amplitude) and phase over time, across different frequencies, and between multiple antennas. In simple terms, it helps us understand how the wireless environment is shaping the signal at any given moment, Figure~\ref{fig:80211a_frame}.

In traditional wireless systems (e.g., Wi-Fi and LTE/5G) which are OFDM-based systems, CSI is represented per subcarrier $H[k]$ where the magnitude and phase represent the attenuation and propagation delay. Accurate CSI estimation is critical as it enables coherent detection, modulation and coding selection, and link adaptation. Traditional systems like Wi-Fi (802.11) use pilot-based channel estimation where frames include known training fields (preambles) that allow the receiver to estimate the channel in the frequency domain on each subcarrier by comparing received pilots with known transmitted pilots.

In static or low-mobility scenarios, this estimate remains accurate throughout the packet duration. However, in higher-mobility environments or for long frames, the channel may vary during transmission.

\subsection{CSI Aging in Traditional Wireless Systems}
\label{sec:csi_aging_traditional}

CSI aging refers to the mismatch that arises when the CSI used for a communication decision does not match the channel at the time that decision is applied. Concretely, if $\hat{H}(t_0)$ is estimated at time $t_0$ but precoding or detection uses this estimate at a later time $t_0 + \Delta t$, mobility and environmental dynamics may cause the true channel $H(t_0 + \Delta t)$ to differ from $\hat{H}(t_0)$. This staleness is driven by channel time-variation (e.g., Doppler effects due to motion) together with estimation, processing, and feedback delays. The impact is particularly pronounced in systems that rely heavily on accurate transmit-side CSI, such as multi-user MIMO, coordinated transmission, and narrow-beam mmWave links.

From a protocol perspective, Wi-Fi provides an intuitive illustration. The receiver estimates the channel using training symbols at the beginning of a packet and applies this estimate to equalize subsequent data symbols. If the channel varies during the packet due to mobility or long frame duration, the equalizer operates with an increasingly outdated channel estimate. In scenarios involving transmit-side adaptation (e.g., beamforming with explicit feedback), additional feedback and scheduling delays further increase the mismatch between the estimated and actual channel. This example highlights that CSI aging is influenced not only by physical channel dynamics but also by protocol timing and system architecture.

To mitigate CSI aging, traditional wireless systems employ several complementary strategies. One approach is frequent channel re-estimation, where pilot symbols are inserted periodically to refresh the channel estimate before significant temporal decorrelation occurs. In high-mobility scenarios, pilot density may be increased to better track rapid channel variations, albeit at the cost of additional overhead. Channel prediction techniques exploit temporal correlation to forecast future channel states using autoregressive models, Kalman filtering, or learning-based predictors. For transmit-side adaptation, particularly in MIMO and multi-user systems, robust precoding and beamforming schemes are designed to explicitly account for CSI uncertainty through statistical or worst-case optimization frameworks. Additionally, user scheduling and link adaptation mechanisms may incorporate CSI reliability, while systems experiencing rapid channel fluctuations may rely on statistical or long-term CSI (e.g., covariance or angular information) that evolves more slowly. At the architectural level, reducing feedback delay, exploiting TDD reciprocity, and shortening frame durations further help limit CSI staleness. Collectively, these approaches allow conventional wireless systems to manage the performance degradation caused by outdated CSI.

\begin{table*}[t]
\centering
\caption{Mitigation Techniques for CSI Aging in Traditional Wireless Systems}
\label{tab:csi_aging_traditional}
\footnotesize
\renewcommand{\arraystretch}{1.15}
\setlength{\tabcolsep}{4pt}
\begin{tabular}{|p{3.0cm}|p{3.7cm}|p{4.8cm}|p{4.8cm}|}
\hline
\textbf{Category} & \textbf{Core Idea} & \textbf{Representative Techniques} & \textbf{Trade-Offs / Limitations} \\
\hline
Frequent Re-Estimation \cite{ma2022channel, li2020channel} & Periodically refresh instantaneous CSI before coherence time is exceeded & Adaptive pilot insertion; mobility-aware training intervals; midamble-based updates & Increased pilot overhead; reduced spectral efficiency; higher energy consumption \\
\hline
Channel Prediction \cite{ma2023channel} & Exploit temporal correlation to forecast future channel realizations & Autoregressive (AR) models; Kalman filtering; Wiener prediction; deep learning-based channel forecasting & Model mismatch under non-stationarity; computational complexity; prediction error accumulation \\
\hline
Robust Transmission Design \cite{zhang2019blind, li2020channel} & Incorporate CSI uncertainty directly into precoder/beamformer design & Stochastic robust optimization; worst-case bounded-error design; outage-constrained beamforming & Conservative performance; reduced peak spectral efficiency \\
\hline
Statistical / Long-Term CSI \cite{ma2022channel, ma2023channel} & Rely on slowly varying channel statistics instead of instantaneous CSI & Covariance-based beamforming; angle-domain precoding; eigenmode transmission & Loss of instantaneous array gain; limited adaptability to fast fading \\
\hline
Protocol-Level Mitigation \cite{li2020channel, qian2021noncoherent} & Reduce delay between CSI acquisition and usage & TDD reciprocity exploitation; compressed feedback; shorter frame duration; fast scheduling & Hardware calibration requirements; signaling complexity; limited scalability \\
\hline
\end{tabular}
\end{table*}

As summarized in Table~\ref{tab:csi_aging_traditional}, existing mitigation strategies span physical-layer design, statistical modeling, and protocol optimization, reflecting the cross-layer nature of CSI aging in conventional wireless systems. Despite these well-developed mitigation strategies in conventional active radios, CSI aging becomes more challenging in backscatter systems due to cascaded channel structures, weaker signal strengths, and more constrained estimation capabilities.

\subsection{Why CSI Aging is More Severe in Backscatter Communications}
\label{sec:csi_aging_severe}

While CSI aging is already recognized as a critical issue in conventional active radio systems, it becomes even more challenging in backscatter communication. The difficulty mainly stems from the underlying architecture and physical limitations of passive tags. Unlike traditional transmitters, backscatter devices do not generate their own RF signals. Instead, they modulate and reflect incident waves from an external illuminator. Because of their strict energy constraints, continuously transmitting pilot symbols for frequent channel re-estimation is generally not feasible. As a result, maintaining up-to-date instantaneous CSI is inherently difficult.

A more subtle difficulty comes from the cascaded structure of the backscatter channel, which fundamentally changes how CSI must be interpreted. In most practical setups, the effective channel is formed by the product of the illuminator-to-tag and tag-to-reader links. This composite structure makes channel estimation considerably more delicate than in traditional wireless systems. This happens because the reader typically observes a strong direct-link signal alongside a much weaker backscattered component. Separating and accurately estimating the cascaded channel under these conditions is nontrivial, and any delay in updating CSI affects both constituent links simultaneously.

The problem is further enhanced by the inherently weak power of the reflected signal, which often necessitates longer observation intervals for reliable detection. In dynamic environments, even moderate mobility of the tag, reader, or nearby objects can cause rapid variations in the cascaded channel, leading to faster temporal decorrelation. Given limited pilot resources, hardware constraints, and persistent direct-path interference, backscatter systems are generally less capable of frequent re-estimation or sophisticated prediction mechanisms compared to conventional single-hop systems.

To further demonstrate the impact of CSI aging on traditional versus backscatter communication systems, consider an 802.11 wireless system with 20 MHz waveform that has a total airtime of approximately 1.4 ms (considering PSDU of 1000 bytes at MCS 0 and code rate 1/2). Since channel estimation is performed using the Long Training Field at the beginning of the packet, the channel estimate must remain valid over an interval of roughly $\Delta t \approx 1.34$ ms (aging window of the WLAN packet in this specific case). 

To characterize the temporal evolution of the wireless channel under mobility, we adopt the classical Clarke-Jakes fading model that describes small-scale fading in rich scattering environments. This model assumes that the received signal is composed of a large number of multipath components arriving from different angles with the transmitter or receiver moving at a constant velocity. Under these assumptions, the time variation of the complex channel coefficient follows a wide-sense stationary random process whose temporal autocorrelation depends on the Doppler frequency $f_D = \frac{v}{c} f_c$. The resulting normalized temporal correlation is given by:

\begin{equation}
\rho(\Delta t) = J_0\!\left(2\pi f_D \Delta t\right)
\label{eq:temporal_correlation}
\end{equation}

where $J_0(\cdot)$ is the zeroth-order Bessel function of the first kind. This expression quantifies how rapidly the channel ``forgets'' its past value as a function of carrier frequency and mobility.

\begin{figure*}[H]
\centering
\includegraphics[width=0.8\textwidth]{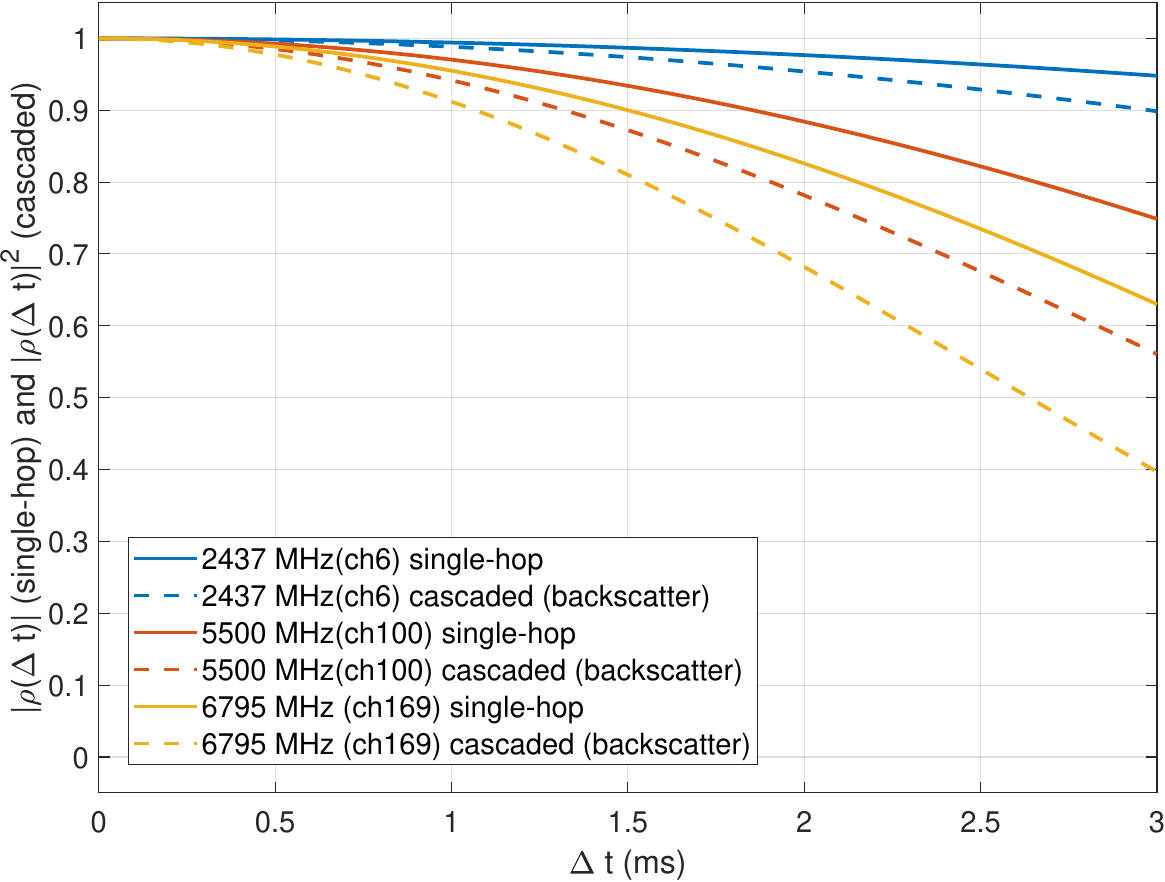}
\caption{Temporal correlation for single-hop (solid) and cascaded backscatter (dashed) links at different carrier frequencies. The backscatter case shows faster decorrelation due to the product of two links.}
\label{fig:temporal_correlation}
\end{figure*}

From Figure~\ref{fig:temporal_correlation}, it is clear that the temporal correlation decreases more rapidly as the carrier frequency increases. This behavior is expected since the Doppler frequency $f_D = \frac{v}{c} f_c$ grows linearly with $f_c$. For indoor mobility, the correlation at 2.4 GHz remains very close to one over millisecond-scale aging windows, indicating that CSI variations within typical packet durations are limited. In contrast, at 5 GHz and 6 GHz, the correlation decreases more noticeably over the same interval, reflecting the stronger sensitivity of higher-frequency systems to motion.

The backscatter case shows an even faster reduction in correlation. Because the effective backscatter channel is formed by the product of the transmitter-tag and tag-receiver links, temporal variations in both hops contribute to the overall channel evolution. As a result, the effective correlation drops more quickly than in the single-hop case. This difference becomes more visible as the aging window increases. Overall, CSI aging is relatively mild for short packet durations at 2.4 GHz under typical indoor motion. However, the effect becomes more significant for longer observation windows, higher carrier frequencies, and cascaded backscatter channels.

In passive backscatter networks, CSI usually refers to reader-side knowledge of the effective backscatter channel or a proxy thereof rather than instantaneous CSI available at the passive tag.

\subsection{Bridge: Connection to Envelope Stability}
\label{sec:csi_envelope_bridge}

The CSI aging phenomena described above directly impact the envelope stability challenges surveyed in Section~\ref{sec:envelope}. Recall that the Backscatter Crest Factor (BCF) quantifies envelope flatness at the tag location:

\begin{equation}
\mathrm{BCF} = \frac{\max_t r(t) - \min_t r(t)}{\mathbb{E}[r(t)]}
\label{eq:bcf_recall}
\end{equation}

When beamforming, spatial filtering, or antenna selection decisions are based on outdated CSI estimates $\hat{H}(t_0)$ rather than the actual channel $H(t_0 + \Delta t)$, the delivered RF envelope becomes less predictable. In flat-fading channels, CSI aging primarily affects the received signal magnitude through scalar channel gain and does not alter normalized temporal envelope metrics such as BCF. However, in frequency-selective environments, where different antenna selections produce different effective channel responses, outdated CSI may lead to suboptimal antenna selection, indirectly affecting the received envelope characteristics and the resulting BCF. Specifically, CSI aging can cause:

\begin{itemize}
    \item \textbf{Increased envelope ripple:} Misaligned beamforming weights may not flatten the spatial power distribution, increasing BCF.
    \item \textbf{Unpredictable nulls and peaks:} Stale CSI may steer energy toward tags that have moved, causing unexpected envelope excursions.
    \item \textbf{Sensing fidelity degradation:} The Sensing Fidelity Index (SFI) introduced in Section~\ref{sec:metrics} depends on phase coherence, which is directly undermined by CSI aging.
\end{itemize}

Thus, the temporal correlation $\rho(\Delta t)$ directly governs how long a channel estimate remains useful for envelope stability. When $\rho(\Delta t)$ drops below a threshold, the BCF can increase significantly, degrading sensing performance. This coupling between temporal dynamics and envelope stability motivates the need for CSI-aware waveform and selection designs.

\subsection{Coherent and Non-Coherent Detection in Backscatter Systems}
\label{sec:coherent_noncoherent}

In any wireless communication system, detection refers to the process of deciding which symbol was transmitted based on the received waveform. Because the wireless channel introduces amplitude attenuation, phase rotation, and transmitter/receiver RF hardware impairments, the receiver must account for these effects before making a reliable decision.

In coherent detection, the receiver explicitly estimates the complex channel response —both magnitude and phase— and compensates for it prior to symbol detection. For instance, in IEEE 802.11 systems, known training symbols such as the Long Training Field (LTF) are used to estimate the frequency-domain channel response on each subcarrier. During data transmission, the received signal is equalized using this complex estimate which corrects amplitude scaling and phase rotation. The equalized symbols are then mapped to the nearest constellation point. When the channel estimate is accurate, coherent detection achieves near-optimal performance.

In backscatter systems, obtaining an accurate complex channel estimate is significantly more challenging. The effective channel is typically the product of the transmitter-to-tag and tag-to-reader links, and the tag does not actively transmit pilot symbols to facilitate estimation. The backscattered signal is often several orders of magnitude weaker than the direct-path component, and these factors make reliable channel estimation difficult, particularly under low SNR conditions.

For this reason, many practical backscatter receivers employ non-coherent detection. In this approach, symbol decisions are made without explicitly estimating or compensating for channel phase. Instead, detection relies on received signal energy, envelope changes, or differential comparisons between successive tag states. Although this avoids the need for explicit CSI estimation and reduces receiver complexity, it generally adds a performance penalty compared to coherent detection and is more sensitive to interference and channel variability.

The performance difference between coherent and non-coherent detection over fading channels is well established in classical communication theory. When binary phase-shift keying (BPSK) over a Rayleigh fading channel is used with coherent detection (where the receiver has perfect knowledge of the complex channel coefficient), the average bit error rate (BER) is given by:

\begin{equation}
P_b^{\text{coherent}} = \frac{1}{2}\left(1 - \sqrt{\frac{\bar{\gamma}}{1+\bar{\gamma}}}\right)
\label{eq:ber_coherent}
\end{equation}

where $\bar{\gamma}$ denotes the average received signal-to-noise ratio (SNR).

In contrast, when differential (i.e., non-coherent) detection is employed without explicit channel phase estimation, the average BER becomes:

\begin{equation}
P_b^{\text{non-coherent}} = \frac{1}{2(1+\bar{\gamma})}
\label{eq:ber_noncoherent}
\end{equation}

These expressions indicate that non-coherent detection adds a performance penalty compared to coherent detection. In particular, at moderate-to-high SNR, coherent detection achieves a steeper BER decay, while non-coherent detection exhibits an approximate 3 dB SNR loss in Rayleigh fading. This performance gap arises because non-coherent detection does not exploit full phase information and is therefore more sensitive to channel fluctuations.

\begin{figure}
\centering
\includegraphics[width=0.49\textwidth]{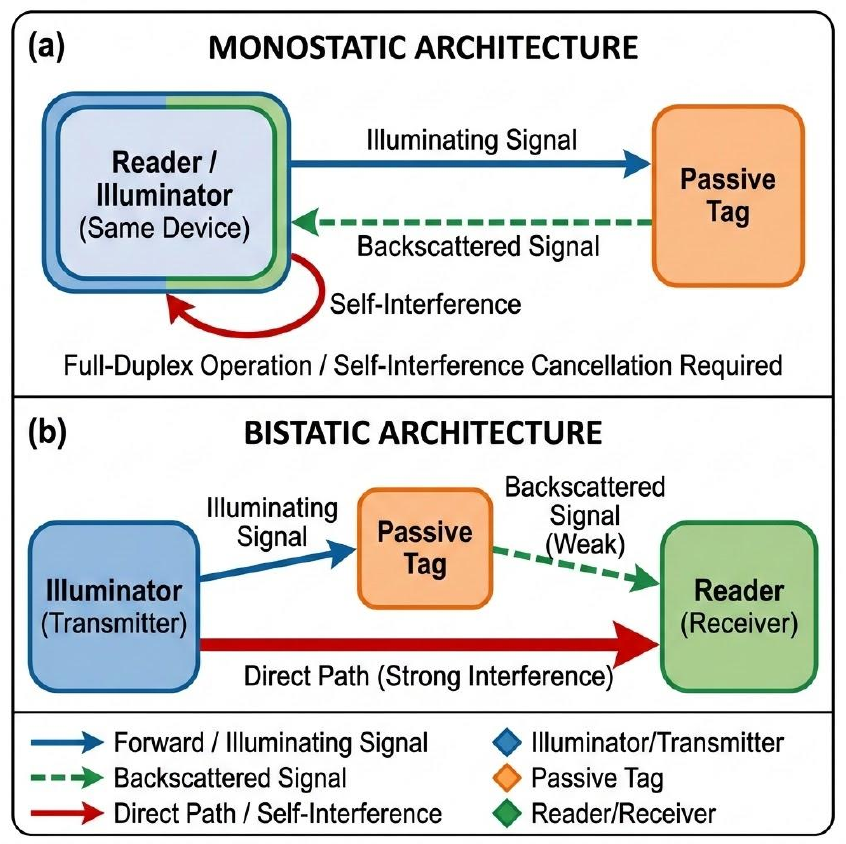}
\caption{Monostatic and bistatic backscatter architectures. In the monostatic configuration, the same device acts as both transmitter and receiver. In the bistatic architecture, the illuminator and reader are separated.}
\label{fig:backscatter_arch}
\end{figure}

Figure~\ref{fig:backscatter_arch} illustrates the two primary backscatter architectures: monostatic and bistatic. In the monostatic configuration, the same device acts as both transmitter and receiver, and the backscattered signal is received at the same location as the illuminating signal. In contrast, the bistatic architecture separates the illuminator and the reader. The reader receives both the strong direct path from the transmitter and the much weaker backscattered component from the tag.

In practical deployments, the bistatic setup offers greater flexibility in geometry and spatial processing, particularly when multiple antennas are available at the reader. However, the presence of a dominant direct path alongside the weak backscattered signal introduces significant detection challenges, especially under non-coherent reception. As transmit power increases, both the direct and backscattered components scale proportionally, and the relative perturbation remains small. Consequently, detection errors become dominated not by thermal noise but by residual interference and channel fluctuations associated with the direct path. In this regime, further increases in SNR do not significantly reduce the error probability, leading to the appearance of a BER floor.

In contrast, if the direct path were perfectly suppressed, the system would revert to a noise-limited fading channel and the BER would continue to decrease with increasing SNR. This highlights the fundamental role of direct-path dominance in creating BER floors in practical backscatter systems.

\subsection{Impact of Multi-Antenna Reception and Spatial Filtering}
\label{sec:multi_antenna}

The limitations of single-antenna non-coherent reception can be alleviated by equipping the reader with multiple antennas. With $M$ receive antennas, the received signal becomes a vector quantity:

\begin{equation}
\mathbf{y}(t) = \mathbf{h}_d x(t) + \mathbf{h}_{bs}(t)\alpha(t) x(t) + \mathbf{n}(t)
\label{eq:multi_antenna_received}
\end{equation}

where $\mathbf{h}_d \in \mathbb{C}^{M \times 1}$ denotes the spatial signature of the direct path and $\mathbf{h}_{bs}(t)$ represents the effective spatial signature of the cascaded backscatter channel.

Since the direct path and the backscattered signal generally arrive from different spatial directions, their corresponding channel vectors are not aligned. This spatial diversity enables the reader to apply linear combining or beamforming:

\begin{equation}
z(t) = \mathbf{w}^H \mathbf{y}(t)
\label{eq:beamforming}
\end{equation}

where $\mathbf{w}$ is a suitably chosen combining vector. By selecting $\mathbf{w}$ to suppress the dominant direct-path component while preserving the backscattered signal, the effective signal-to-clutter ratio can be significantly improved.

In contrast to single-antenna reception, where the direct path directly limits detection performance, multi-antenna spatial filtering reduces residual interference prior to non-coherent detection. As a result, the system transitions from an interference-limited regime toward a noise-limited regime, thereby lowering the BER floor. The achievable gain depends on the angular separation between the direct and backscattered components, the number of antennas, and the stability of the spatial channel.

\begin{table*}[t]
\centering
\caption{CSI Assumptions and Spatial Processing in Multi-Antenna Non-Coherent Backscatter Receivers}
\label{tab:csi_spatial}
\footnotesize
\renewcommand{\arraystretch}{1.1}
\setlength{\tabcolsep}{4pt}
\begin{tabular}{|p{6.0cm}|p{2.8cm}|p{2.5cm}|p{2.0cm}|p{2.5cm}|}
\hline
\textbf{Architecture / Reference} & \textbf{Spatial Method} & \textbf{CSI Type} & \textbf{Fading Model} & \textbf{Key Gain} \\
\hline
Ambient \cite{QianTWC2017} & Baseline non-coherent & None & Block & BER vs RCD \\
\hline
Ambient \cite{DuanSPAWC2018} & OFDM covariance distance & None & Block & 1 OFDM symbol detection \\
\hline
Ambient \cite{DevineniGC2020} & AoA tracking + DL cancel & AoA only & Time-selective & $\sim$9 dB antenna gain \\
\hline
Bistatic \cite{WangGC2021} & Single-beamformer & Statistical & Quasi-static & 1-3.6 dB from optimal \\
\hline
Ambient \cite{YigitlerTWC2023} & MAP projection beamforming & Correlation matrices & Slow & Near-optimal performance \\
\hline
Bistatic \cite{KaplanGC2022} & Null-space DL suppression & DL CSI & Static & Dynamic range reduction \\
\hline
Ambient (FDA) \cite{HuTWC2022} & Spatio-frequency separation & Structural & Block & Improved DL robustness \\
\hline
\end{tabular}
\end{table*}

Table~\ref{tab:csi_spatial} summarizes multi-antenna and AoA-driven techniques for non-coherent backscatter reception, with emphasis on their CSI assumptions and channel modeling. Several observations can be drawn. First, most spatial suppression methods rely on at least partial knowledge of the direct link (DL), such as AoA estimates, statistical correlation matrices, or explicit DL CSI. Even in non-coherent frameworks, some form of spatial signature information is typically required to enable direct-path cancellation or beamforming. The majority of the reported results assume block or quasi-static fading, where spatial signatures remain stable over the detection interval. Only limited works explicitly model time-selective fading, and even fewer analyze robustness under rapid channel variation. This indicates that many multi-antenna backscatter receivers implicitly depend on slowly varying spatial channels for effective interference suppression.

\subsection{Limitations Under High Mobility and Short Coherence Time}
\label{sec:mobility_limitations}

Most non-coherent backscatter receivers are analyzed assuming that the channel remains approximately constant over the duration of a detection window, corresponding to a block-fading assumption. Under higher mobility, the coherence time decreases and the channel may evolve within a single symbol interval or averaging window. For energy-based detectors, the decision statistic is typically formed by averaging the received signal energy over multiple samples. If the channel gain fluctuates within that interval, the samples are no longer drawn from the same distribution, and the averaging operation does not provide the intended variance reduction. As a result, the separation between the two tag states becomes less distinct, even if the average SNR remains unchanged.

Differential detection schemes also rely on short-term channel stability with the assumption that the channel phase does not vary significantly between adjacent symbols. Under higher Doppler conditions, this assumption weakens and the differential phase reference becomes noisy. This introduces additional decision errors beyond those caused by thermal noise. Multi-antenna receivers face a similar issue where spatial filtering techniques such as AoA-based nulling assume that spatial signatures remain stable over the beamforming interval. When mobility induces small angular or multipath changes, previously designed nulls may no longer align perfectly with the direct path. Even minor misalignment can noticeably reduce direct-path suppression and increase residual interference.

As the effective channel in backscatter systems is cascaded, the composite channel can decorrelate faster than a conventional single-hop channel. This increases sensitivity to mobility compared to traditional wireless links.

\subsection{Tag-Centric Metrics Under CSI Aging}
\label{sec:tag_centric_metrics}

\begin{figure}
\centering
\includegraphics[width=0.5\textwidth]{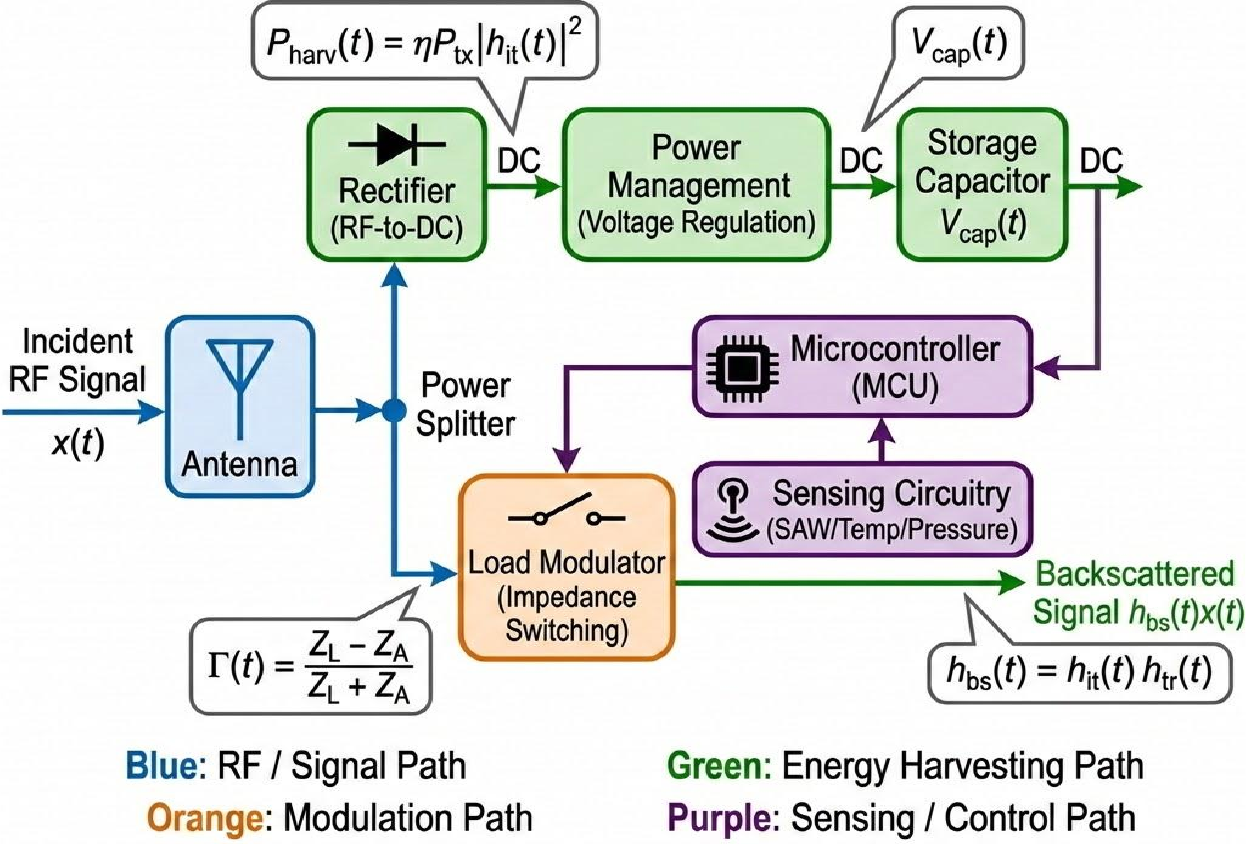}
\caption{Passive backscatter tag architecture. The incident RF signal is split between the energy harvesting path and the modulation path, where the load impedance is switched to encode information in the reflected signal.}
\label{fig:tag_arch}
\end{figure}

Figure~\ref{fig:tag_arch} illustrates a simplified passive backscatter tag architecture. Unlike active radios, the tag does not generate its own carrier and relies entirely on the incident RF signal for both energy harvesting and communication. The antenna captures the incoming waveform and interfaces with two functional paths: the energy harvesting path and the modulation path. The rectifier converts the RF signal into DC power providing harvested energy:

\begin{equation}
P_{\text{harv}}(t) = \eta P_{\text{tx}} |h_{it}(t)|^2
\label{eq:harvested_power}
\end{equation}

where $h_{it}(t)$ denotes the illuminator-to-tag channel and $\eta$ represents rectifier efficiency. This harvested energy charges a storage capacitor whose voltage $V_{\text{cap}}(t)$ determines whether the microcontroller and sensing circuitry can operate reliably. Simultaneously, the load impedance $Z_L$ is switched between discrete states, altering the antenna reflection coefficient $\Gamma = \frac{Z_L - Z_A}{Z_L + Z_A}$ and thereby modulating the amplitude (and sometimes phase) of the reflected signal. The backscattered waveform observed at the reader depends on the cascaded channel $h_{bs}(t) = h_{it}(t) h_{tr}(t)$, where $h_{tr}(t)$ is the tag-to-reader channel.

From this architecture, several tag-centric performance metrics naturally emerge. First, \textbf{harvested energy stability} determines whether the tag can sustain continuous operation. Variations in $|h_{it}(t)|^2$ due to fading or beam misalignment directly translate into fluctuations in harvested power and capacitor voltage, potentially causing brownout events or reduced sensing duty cycles. Second, \textbf{modulation detection reliability} depends on the effective signal-to-clutter ratio at the reader:

\begin{equation}
\text{SCR}(t) = \frac{|h_{it}(t) h_{tr}(t)|^2}{|h_d(t)|^2 + \sigma^2}
\label{eq:scr}
\end{equation}

where $h_d(t)$ represents the direct-path channel. Since the backscatter link is cascaded, small temporal variations in either hop can significantly alter the modulation contrast and increase bit error probability. Third, in sensing-aware tags, \textbf{sensing fidelity} relies on the stability of the received envelope at the detector circuit. Any drift in the baseline channel conditions can corrupt feature extraction, reduce detection sensitivity, or introduce false sensing events.

Although the passive tag itself does not perform channel estimation, CSI aging at the system level indirectly affects all of these metrics. When beamforming, spatial filtering, or tag selection decisions are based on outdated channel estimates $\hat{h}(t_0)$ rather than the actual channel $h(t_0+\Delta t)$, the delivered RF energy, effective cascaded gain, and interference suppression may become suboptimal. Under temporal correlation $\rho(\Delta t)$, decreasing correlation reduces both the predictability of harvested power and the effectiveness of direct-path suppression, thereby increasing energy instability, raising BER, and degrading sensing performance. Consequently, CSI aging propagates through the hardware blocks of the passive tag and manifests as tangible degradations in tag-centric operation.

\subsection{Bridge: Connection to Near-Far Mitigation }
\label{sec:csi_nearfar_bridge}

CSI aging also interacts critically with the near-far mitigation techniques surveyed in Section~\ref{sec:nearfar}. Recall that transmit antenna selection (TAS) is a low-complexity spatial mitigation technique that selects the antenna minimizing the Backscatter Crest Factor (BCF) at tags. It is important to note that in flat-fading channels, antenna selection changes only the received signal magnitude and does not alter normalized BCF. The BCF-related effects discussed here are relevant in frequency-selective propagation, where different antenna selections may modify the effective channel response and the received envelope characteristics.

However, TAS effectiveness depends on the currency of CSI:

\begin{itemize}
    \item \textbf{Outdated TAS decisions:} When TAS is performed using stale CSI $\hat{\mathbf{h}}(t_0)$ instead of $\mathbf{h}(t_0+\Delta t)$, the selected antenna may no longer provide optimal spatial nulling (for near-field tags) or energy steering (for far-field tags).
    \item \textbf{Increased saturation/starvation:} A TAS decision that balanced incident power at $t_0$ may, after channel decorrelation, increase both $\eta_{\mathrm{sat}}$ and $\eta_{\mathrm{starve}}$ (the fractions of saturated and starved tags defined in Section~\ref{sec:nearfar}).
    \item \textbf{Adaptation overhead tradeoff:} Frequent TAS updates improve tracking but increase CSI acquisition overhead—a critical constraint in passive networks.
\end{itemize}

Similarly, beamforming and IRS-assisted spatial redistribution rely on CSI that can age rapidly in mobile environments. The temporal correlation $\rho(\Delta t)$ from Equation~\ref{eq:temporal_correlation} directly governs how long spatial configurations remain effective.

\subsection{CSI Acquisition and Channel Knowledge in Passive Backscatter Networks}
\label{sec:csi_acquisition}

Table~\ref{tab:csi_aging_passive_backcom} summarizes representative works on channel knowledge acquisition and CSI aging in passive backscatter networks. A first observation is that the term ``CSI'' is used more broadly in passive backscatter than in conventional active radios. In many passive BackCom systems, the tag does not estimate and feed back a full complex channel. Instead, the reader acquires a proxy of the effective link, such as received signal strength, an effective SNR, a cascaded channel coefficient, or a spatial feature such as angle-of-arrival. Therefore, in this survey, CSI aging should be interpreted as the temporal mismatch between the channel knowledge acquired during training and the true effective backscatter channel during payload transmission.

The works in Table~\ref{tab:csi_aging_passive_backcom} can be divided into three broad classes. The first class consists of \emph{selection-based outage analyses}, where the reader ranks tags using a scalar quality metric and then suffers performance loss when that metric becomes outdated or imperfect. These papers are particularly important because they show that, in passive systems, the quantity called ``CSI'' is often only a reader-side channel-quality indicator rather than a full channel vector.

The second class consists of \emph{explicit channel-estimation papers}, where the reader estimates either the cascaded backscatter channel or both the direct and cascaded links using dedicated training. These papers are central to the CSI-aging discussion because they define how the channel knowledge is obtained in the first place, what overhead is required, and which estimates are likely to become stale when the channel varies over time.

The third class consists of \emph{reduced-CSI or CSI-robust designs}, which arise when accurate tracking of instantaneous backscatter channels is too costly or unreliable. These studies are useful because they show that ``passive CSI'' is not impossible, but it is usually reader-centric, architecture-dependent, and strongly constrained by training overhead and hardware simplicity.

From a survey perspective, the main issue is not only \emph{whether} CSI exists, but also \emph{what form} it takes, \emph{where} it is estimated, and \emph{how} it degrades with time. In passive backscatter, aging may arise from delay between training and data phases, mobility of the reader/tag/ambient source, oscillator drift, intermittent tag activation, hardware impairments, pilot contamination, or uncertainty in the reflection state. Consequently, the performance impact of aging depends on the role of the estimate: stale RSS-based metrics mainly cause suboptimal tag selection, stale cascaded-channel estimates degrade coherent combining and beamforming, and rapid fading often motivates non-coherent or pilot-assisted alternatives.

\begin{table*}[t]
\centering
\caption{CSI/CSI-proxy estimation and CSI aging in passive backscatter networks}
\label{tab:csi_aging_passive_backcom}

\small
\renewcommand{\arraystretch}{1.1}
\setlength{\tabcolsep}{4pt}

\begin{tabular}{|
>{\raggedright\arraybackslash}p{4.8cm}|
>{\raggedright\arraybackslash}p{3.3cm}|
>{\raggedright\arraybackslash}p{3.5cm}|
>{\raggedright\arraybackslash}p{4.5cm}|}
\hline
\textbf{Scenario / Reference} &
\textbf{Estimated Quantity} &
\textbf{Impairment Model} &
\textbf{Relevance to CSI Aging} \\
\hline

Monostatic reciprocal multi-tag \cite{li2020channel} &
Reader-side tag quality metric ($|h_n|^2$) &
Explicit outdated CSI with temporal correlation $\rho$ &
Power-based CSI proxy; selection gains degrade with aging \\
\hline

Multi-tag ambient backscatter \cite{kim2021channel} &
Tag-selection metric at reader &
Outdated CSI with RF hardware impairments &
Ambient BackCom with stale selection and hardware effects \\
\hline

Tag-selection BackCom \cite{chen2022antenna} &
Selection metric based on estimated channel quality &
Channel estimation error &
Decouples delay aging from estimation and mismatch effects \\
\hline

Monostatic multi-tag \cite{vicario2022antenna} &
Order-statistic-based tag selection &
No explicit aging model &
Baseline for tag selection without delay or estimation errors \\
\hline

Monostatic full-duplex MIMO reader \cite{ma2022channel} &
Reciprocal forward/backward channel via LS/LMMSE &
CE-resource tradeoff; no aging model &
Clear pipeline for effective channel estimation \\
\hline

Multi-tag MIMO BackCom \cite{nguyen2022mimo} &
Multi-tag channels at MIMO reader &
CE under passive hardware constraints &
Channel acquisition scaling with many passive tags \\
\hline

General BackCom networks \cite{kim2023direct} &
Direct and cascaded channels &
Pilot contamination; training design &
Bridges classic pilot schemes to one-shot multi-tag estimation \\
\hline

Ambient BackCom with massive-array reader \cite{zhang2019blind} &
Channel gains and DoAs &
Spatial estimation; no delay model &
Relevant when CSI includes spatial features \\
\hline

Practical BackCom with activation threshold \cite{li2020channel} &
Channel estimates with unknown BD activation &
Circuit sensitivity constraint &
CSI uncertainty from activation, not only aging \\
\hline

Ambient BackCom (fast fading) \cite{DevineniGC2020} &
Noncoherent statistics (vs.\ CSI) &
AR(1) time-selective fading &
Supports avoiding full CSI tracking in short coherence \\
\hline

Field demo with LTE pilots \cite{kim2021channel} &
Stable pilot-derived features &
Traffic variability; sync issues &
Tracking stable pilots over bursty ambient signals \\
\hline

Passive tag-to-tag network \cite{griffin2022multistatic} &
Amplitude and phase of tag-to-tag channel &
No aging model; multiphase probing &
Passive channel estimation via specialized protocols \\
\hline

Monostatic multiantenna BackCom \cite{boyer2012space} &
Backscatter and forward channels &
Model mismatch; reduced pilot overhead &
Data-driven estimation replacing exact-statistics models \\
\hline

\end{tabular}
\end{table*}

\subsection{Open Research Challenges}
\label{sec:csi_open_challenges}

Several open challenges remain in addressing CSI aging for sensing-aware backscatter networks:

\begin{itemize}
    \item \textbf{Low-power, tag-friendly detection methods:} Designing detection methods robust to rapid channel changes without requiring frequent pilot transmission from passive tags.
    
    \item \textbf{Spatial diversity and coding strategies:} Integrating spatial diversity and coding strategies to improve BER under fast fading conditions in cascaded backscatter channels.
    
    \item \textbf{CSI-aware scheduling and selection:} Developing CSI-aware scheduling or selection frameworks for multiple backscatter tags sharing ambient signals, accounting for CSI aging in decision metrics.
    
    \item \textbf{Blind estimation under mobility:} Extending blind estimation techniques to high-mobility scenarios with minimal pilot or feedback overhead.
    
    \item \textbf{Joint temporal-spatial optimization:} Formulating optimization frameworks that jointly consider CSI aging (temporal) and near-far mitigation (spatial) to balance overhead, accuracy, and sensing fidelity.
    
    \item \textbf{Learning-based CSI prediction:} Exploring machine learning approaches (e.g., transformers, recurrent neural networks) for predicting channel evolution in cascaded backscatter links, building on the temporal correlation model in Equation~\ref{eq:temporal_correlation}.
\end{itemize}

\subsection{Summary}
\label{sec:csi_summary}

This section has formalized the CSI aging challenge in sensing-aware backscatter communications. We introduced temporal correlation models for cascaded backscatter channels, compared coherent and non-coherent detection performance, surveyed multi-antenna spatial filtering techniques, and analyzed tag-centric metrics under CSI aging. The key insight is that backscatter systems are uniquely vulnerable to CSI aging due to their cascaded channel structure, lack of active pilots, and direct-path interference. Furthermore, CSI aging couples with both envelope stability (Section~\ref{sec:envelope}) and near-far mitigation (Section~\ref{sec:nearfar}), creating cross-layer dependencies that require joint optimization. Addressing these coupled challenges remains an important direction for future research in sensing-aware backscatter communications.

\section{Conclusions and Future Directions}
\label{sec:conclusion}

\subsection{Summary of Contributions}
\label{sec:conclusion_summary}

This survey has presented a comprehensive review of sensing-aware backscatter communications, with a focus on three tightly coupled challenges: envelope stability, the near-far interference gap, and CSI aging. Unlike existing surveys that adopt a communication-centric perspective, this work has taken a tag-centric viewpoint, emphasizing how physical-layer design choices impact both communication reliability and sensing fidelity.



The key contributions of this survey are:

\begin{itemize}
    \item \textbf{Envelope stability (Section~\ref{sec:envelope}):} We formalized the concept of envelope stability through tag-centric metrics including the Backscatter Crest Factor (BCF), Envelope Stability Factor (ESF), and Sensing Fidelity Index (SFI). We re-evaluated classical PAPR reduction techniques through a sensing-aware lens and identified that no existing technique simultaneously controls peaks, nulls, and ripple—the three degradation channels for passive sensing.
    
    \item \textbf{Near-far interference gap (Section~\ref{sec:nearfar}):} We quantified the severity of near-far disparity in backscatter systems through the near-far power ratio $\Gamma_{\mathrm{NF}}$ and surveyed six spatial mitigation techniques. The comparative analysis revealed that no single technique fully resolves the simultaneous challenges of energy starvation and rectifier saturation in dense deployments.
    
    \item \textbf{CSI aging (Section~\ref{sec:csi_aging}):} We characterized temporal correlation in cascaded backscatter channels using the Clarke-Jakes model and demonstrated that backscatter links decorrelate faster than conventional single-hop links. We surveyed coherent and non-coherent detection methods, multi-antenna spatial filtering, and tag-centric metrics under CSI aging.
\end{itemize}

\subsection{Key Takeaways}
\label{sec:conclusion_takeaways}

Table~\ref{tab:key_takeaways} summarizes the three core challenges, their root causes, primary metrics, and representative solutions surveyed in this work.

\begin{table}[H]
\centering
\caption{Summary of Core Challenges in Sensing-Aware Backscatter Communications}
\label{tab:key_takeaways}
\footnotesize
\renewcommand{\arraystretch}{1.2}
\setlength{\tabcolsep}{3pt}
\begin{tabular}{|p{1.4cm}|p{2.2cm}|p{1.2cm}|p{3.0cm}|}
\hline
\textbf{Challenge} & \textbf{Root Cause} & \textbf{Primary Metrics} & \textbf{Representative Solutions} \\
\hline
Envelope Stability & High-PAPR multi-carrier illumination & BCF, ESF, SFI & PAPR Reduction (Clipping, SLM, PTS), MD-OFDM, Multisine WPT waveforms \\
\hline
Near-Far Gap & Double path-loss propagation and spatial power imbalance & $\Gamma_{\mathrm{NF}}$, $\eta_{\mathrm{starve}}$, $\eta_{\mathrm{sat}}$ & TAS, Beamforming, IRS, Cooperative Relaying, Multi-Reader Diversity \\
\hline
CSI Aging & Cascaded channel dynamics & $T_c$, $\rho(\Delta t)$, SFI & Kalman Filtering, Non-Coherent Detection, AoA-based Spatial Filtering \\
\hline
\end{tabular}
\end{table}

\subsection{Synthesis: The Interconnected Nature of the Three Challenges}
\label{sec:conclusion_synthesis}

A central insight of this survey is that the three challenges—envelope stability, near-far mitigation, and CSI aging—are not independent but deeply interconnected:

\begin{itemize}
    \item \textbf{Envelope stability $\leftrightarrow$ Near-far:} The spatial power imbalance characterized by the near-far ratio directly determines whether tags experience envelope fluctuations that drive them into saturation or starvation. Spatial mitigation techniques such as TAS aim to flatten the envelope spatially, directly impacting BCF.
    
    \item \textbf{Near-far $\leftrightarrow$ CSI aging:} The effectiveness of spatial mitigation techniques depends critically on the currency of CSI. When TAS or beamforming decisions are based on outdated estimates, the selected configuration may no longer balance incident power, exacerbating near-far disparity.
    
    \item \textbf{CSI aging $\leftrightarrow$ Envelope stability:} Stale CSI leads to misaligned beamforming or selection weights, causing unpredictable envelope fluctuations that increase BCF and degrade sensing fidelity (SFI).
\end{itemize}

This interconnectedness implies that optimizing one dimension in isolation is insufficient. A holistic approach—one that jointly considers waveform design, spatial diversity, and temporal tracking—is required for robust sensing-aware backscatter operation. Table~\ref{tab:key_takeaways} summarizes the core challenges and their relationships.

Viewed collectively, the three challenges can be interpreted as different manifestations of the same underlying problem: maintaining reliable sensing operation under imperfect illumination. Envelope stability governs the quality of the incident excitation, near-far effects determine how that excitation is distributed across space, and CSI aging determines how accurately the network can adapt to changes over time. Together, these dimensions define the operating envelope of sensing-aware backscatter systems.

\subsection{Overarching Open Challenges}
\label{sec:conclusion_challenges}

Beyond the specific open challenges identified within each section, several overarching challenges cut across all three dimensions:

\begin{itemize}
    \item \textbf{Cross-layer optimization:} Waveform shaping (envelope stability), antenna selection (near-far), and channel tracking (CSI aging) are typically optimized separately. Developing joint optimization frameworks that simultaneously consider all three dimensions remains an open problem.
    
    \item \textbf{Hardware-aware design:} Power amplifier nonlinearity at the illuminator and rectifier nonlinearity at the tag interact in non-trivial ways with envelope fluctuations, spatial selection, and CSI aging. A unified hardware-aware modeling framework is currently lacking.
    
    \item \textbf{Standards compatibility:} Ambient IoT standards (802.11bp, BLE Amb-IoT) impose constraints on illuminator behavior. Designing sensing-aware backscatter systems that operate within these constraints while maintaining envelope stability and sensing fidelity is an ongoing challenge.
    
    \item \textbf{Scalability in dense deployments:} As tag density increases, the probability that some tags lie outside the acceptable dynamic range grows. Characterizing scaling laws for envelope stability, near-far mitigation, and CSI tracking in dense passive networks remains an open theoretical question.
    
    \item \textbf{Real-world deployment:} Industrial environments with moving robots, forklifts, and human operators create dynamic conditions that challenge all three dimensions simultaneously. Validating proposed techniques in real-world settings is essential but under-explored.
\end{itemize}

\subsection{Future Research Directions}
\label{sec:conclusion_future}

Looking ahead, several promising research directions emerge from this survey:

\begin{itemize}
    \item \textbf{Machine learning for joint optimization:} The interconnected nature of the three challenges makes them well-suited for machine learning approaches. Deep reinforcement learning could jointly optimize waveform shaping, antenna selection, and channel tracking. Transformer-based architectures show promise for CSI prediction in cascaded backscatter channels. These techniques can learn complex spatial-temporal dependencies that are difficult to model analytically.
    
    \item \textbf{Integrated sensing and communication (ISAC) co-design:} The evolution toward ISAC creates new opportunities for sensing-aware backscatter. Waveform design that jointly optimizes communication rate and sensing fidelity, while maintaining envelope stability, represents a rich area for future work. The metrics introduced in this survey—BCF, ESF, SFI—provide a foundation for such multi-objective optimization.
    
    \item \textbf{Reconfigurable intelligent surfaces (RIS) for sensing-aware environments:} RIS technology offers new degrees of freedom for spatial mitigation and envelope shaping. Future RIS architectures could actively shape the incident RF envelope to minimize BCF across heterogeneous tag locations, effectively acting as a ``spatial envelope flattener.'' Integrating RIS with envelope-aware waveform design and CSI tracking is an emerging area with significant potential.
    
    \item \textbf{Digital twin-assisted design and optimization:} Digital twins—software replicas of physical environments—can pre-calculate optimal antenna selections, waveform parameters, and CSI tracking strategies before deployment. In dynamic warehouse environments, digital twins could continuously update recommendations based on real-time sensor data, enabling proactive rather than reactive mitigation of the three challenges.
    
    \item \textbf{6G Ambient IoT ecosystems:} Emerging standardization efforts such as IEEE 802.11bp for Ambient Power Communication, 3GPP Ambient IoT studies, BLE-based ambient/backscatter proposals, and Wi-Fi sensing standardization under IEEE 802.11bf collectively motivate coexistence between communication, sensing, and ultra-low-power devices. Designing sensing-aware systems that operate within these standards while addressing the three challenges identified in this survey is the ultimate research frontier.
\end{itemize}

\subsection{Concluding Remarks}
\label{sec:conclusion_remarks}

The vision of ubiquitous, battery-free sensing networks has driven backscatter communication from a niche technology to a foundational enabler for 6G IoT. However, realizing this vision requires more than low-power communication—it requires sensing-aware design that accounts for the fundamental constraints of passive hardware.

This survey has formalized the \textbf{``Illuminator's Dilemma''} that lies at the heart of sensing-aware backscatter: the conflict between high-efficiency communication waveforms and the stringent dynamic range requirements of passive tags. Through the lenses of envelope stability, near-far mitigation, and CSI aging, we have shown that sensing fidelity depends not on any single dimension but on their joint optimization.

Ultimately, the future of battery-free sensing will not be determined solely by advances in communication theory, waveform design, or antenna technologies in isolation. Rather, it will depend on the ability to \emph{jointly engineer illumination, propagation, and sensing as a unified system}. By framing envelope stability, near-far mitigation, and CSI aging as interconnected manifestations of the broader Illuminator's Dilemma, this survey provides a foundation for the next generation of sensing-aware backscatter networks and highlights the path toward scalable, intelligent, and ubiquitous ambient sensing in 6G and beyond.

The Backscatter Crest Factor (BCF), Envelope Stability Factor (ESF), and Sensing Fidelity Index (SFI) introduced in this survey provide a unified framework for evaluating and comparing sensing-aware backscatter systems. We hope this survey serves as a foundation for future research that bridges waveform design, spatial diversity, and temporal tracking to enable truly robust, battery-free sensing networks for the next decade and beyond.

\printcredits

\bibliographystyle{cas-model2-names}

\bibliography{main}





\end{document}